\documentclass[11pt,a4paper]{JHEP3}
\pdfoutput=1

\preprint{Cavendish-HEP-12/07\\CERN-PH-TH/2012-099\\MCNET-12-04}

\usepackage{cite}
\usepackage{amssymb}
\usepackage{amsmath}
\usepackage{amsfonts}
\usepackage{enumerate}
\usepackage{slashed}
\usepackage[pdftex]{graphicx}
\newcommand{\beq}{\begin{equation}}
\newcommand{\eeq}{\end{equation}}
\newcommand{\beqn}{\begin{eqnarray}}
\newcommand{\eeqn}{\end{eqnarray}}
\newcommand{\bmat}{\begin{pmatrix}}
\newcommand{\emat}{\end{pmatrix}}

\def\as{\alpha_\mathrm{S}}
\def\eq#1{Eq.~(\ref{#1})}

\def\fig#1{Fig.~\ref{#1}}
\def\figs#1{Figs.~\ref{#1}}
\def\tab#1{Tab.~\ref{#1}}

\def\sec#1{Section~\ref{#1}}
\def\app#1{Appendix~\ref{#1}}
\newcommand{\Sherpa}{S\protect\scalebox{0.77}{HERPA}}
\newcommand{\Pythia}{P\protect\scalebox{0.77}{YTHIA}}
\newcommand{\Herwig}{H\protect\scalebox{0.77}{ERWIG}}
\newcommand{\Css}{CS\protect\scalebox{0.77}{SHOWER}}

\newcommand{\MCNLO}{MC\protect\scalebox{0.77}{@NLO}}
\newcommand{\POWHEG}{P\protect\scalebox{0.77}{OWHEG}}

\title{QCD Coherence and the Top Quark Asymmetry}

\author{Peter Skands$^1$, Bryan Webber$^{2}$ and Jan Winter$^1$\\
$^1$CERN PH-TH, Geneva 23, CH-1211 Switzerland \\
$^2$Cavendish Laboratory, University of Cambridge, JJ Thomson Avenue,  Cambridge, UK\\
Email: \email{peter.skands@cern.ch, webber@hep.phy.cam.ac.uk, jwinter@cern.ch}}

\abstract{
Coherent QCD radiation in the hadroproduction of top quark pairs leads
to a forward--backward asymmetry that grows more negative with
increasing transverse momentum of the pair.  This feature is present in
Monte Carlo event generators with coherent parton showering, even
though the production process is treated at leading order and
has no intrinsic asymmetry before showering.  In addition, depending
on the treatment of recoils, showering can produce a positive
contribution to the inclusive asymmetry.
We explain the origin of these features, compare them in
fixed-order calculations and the \Herwig++, \Pythia\ and \Sherpa\
event generators, and discuss their implications.
}   


\begin{document}   
\section{Introduction}
The observation of a substantial forward--backward asymmetry in
the production of top quark pairs at the
Tevatron~\cite{Arbazov:2007qb,Aaltonen:2008hc,Aaltonen:2011kc,Abazov:2011rq,Aaltonen:2012}
has prompted renewed theoretical study of Standard Model predictions
for this quantity: see Refs.~\cite{Almeida:2008ug,Dittmaier:2008uj,Kidonakis:2011zn,Ahrens:2011uf,Hollik:2011ps,Kuhn:2011ri,Alioli:2011as,Melnikov:2011qx,Manohar:2012rs,Campbell:2012uf}
and references therein.  Quantities of particular interest are the
distributions of the asymmetry with respect to some observable,
generally denoted by $O$\/:
\beq\label{eq:AFBdef}
A_\mathrm{FB}(O)\;=\;
\frac{\;\left.\dfrac{d\sigma}{dO}\right\rfloor_{\Delta y>0}\,-\,
        \left.\dfrac{d\sigma}{dO}\right\rfloor_{\Delta y<0}\;}
     {\left.\dfrac{d\sigma}{dO}\right\rfloor_{\Delta y>0}\,+\,
      \left.\dfrac{d\sigma}{dO}\right\rfloor_{\Delta y<0}}\ ,
\eeq
where the rapidity difference is defined as $\Delta y=y_t-y_{\bar t}$.
Examples of possible observables $O$\/ are the 
invariant mass $m_{t\bar t}$ and transverse momentum $p_{T,t\bar t}$
of the pair.  For the former, QCD predicts an asymmetry that increases
with $m_{t\bar t}$~\cite{Almeida:2008ug,Ahrens:2011uf}.  For the latter, the
prediction changes
sign as $p_{T,t\bar t}$ increases, since the NLO loop contribution is
positive at $p_{T,t\bar t}=0$ while the real-emission contribution at
$p_{T,t\bar t}>0$ is negative.  After matching to parton showers using
the \MCNLO\ prescription~\cite{Frixione:2002ik,Frixione:2003ei}, the predicted
cross-over is around $p_{T,t\bar t}\approx 25$ GeV: see \fig{fig:D0_pt_asy}.

\FIGURE[!t]{
  \centering
  \includegraphics[scale=0.67]{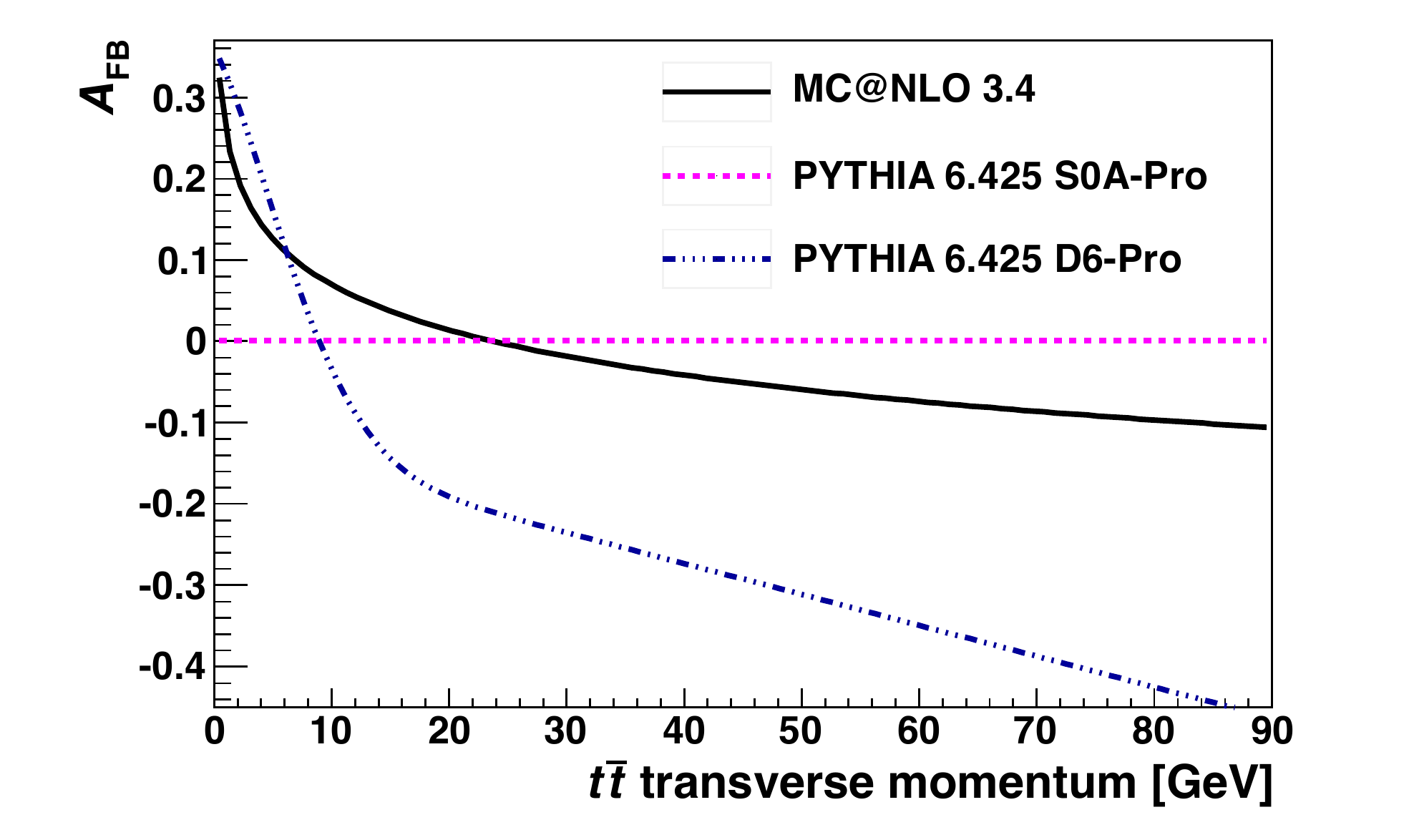}
  \vspace*{-2mm}
  \caption{\label{fig:D0_pt_asy}%
    Top quark forward--backward asymmetry at the Tevatron, 
    as a function of the
    transverse momentum of the pair: results from \MCNLO\ and \Pythia.
    Figure from Ref.~\cite{Abazov:2011rq}.}
}

A surprising fact, also shown in \fig{fig:D0_pt_asy}, is that a
leading-order parton shower event generator such as \Pythia, with
appropriate settings, displays a qualitatively similar $p_T$-dependent
asymmetry, even though the LO production processes have no
asymmetry.  As we shall see, the same is true of the \Herwig++\ and
\Sherpa\ event generators, although their quantitative predictions
differ.\footnote{For a recent review of Monte Carlo event generators,
  see \cite{Buckley:2011ms}.}

The explanation is a nice illustration of the QCD coherence of parton
showering.  It is true that the inclusive asymmetry built into the LO
generators is zero.  However, in the hard process  $q\bar q\to t\bar
t$\/ the colour flows from the incoming quark to the top quark and from
the antiquark to the antitop lead to a more violent acceleration of
colour, and consequently more QCD radiation, when the top is produced
backwards in the $q\bar q$\/ frame than when it goes forwards,
as illustrated in \fig{fig:ttbar_coh}.
The additional radiation when the top goes backwards pushes the
recoiling pair to higher transverse momentum.  Correspondingly, events
with forward-moving tops are left at lower transverse momentum,
leading to the behaviour seen in  \fig{fig:D0_pt_asy}.
The effect vanishes at threshold and becomes
more and more marked as the invariant mass of the pair increases, due
to the increasing amount and scale of QCD radiation.

\FIGURE[!t]{
  \centering
  \includegraphics[scale=0.5]{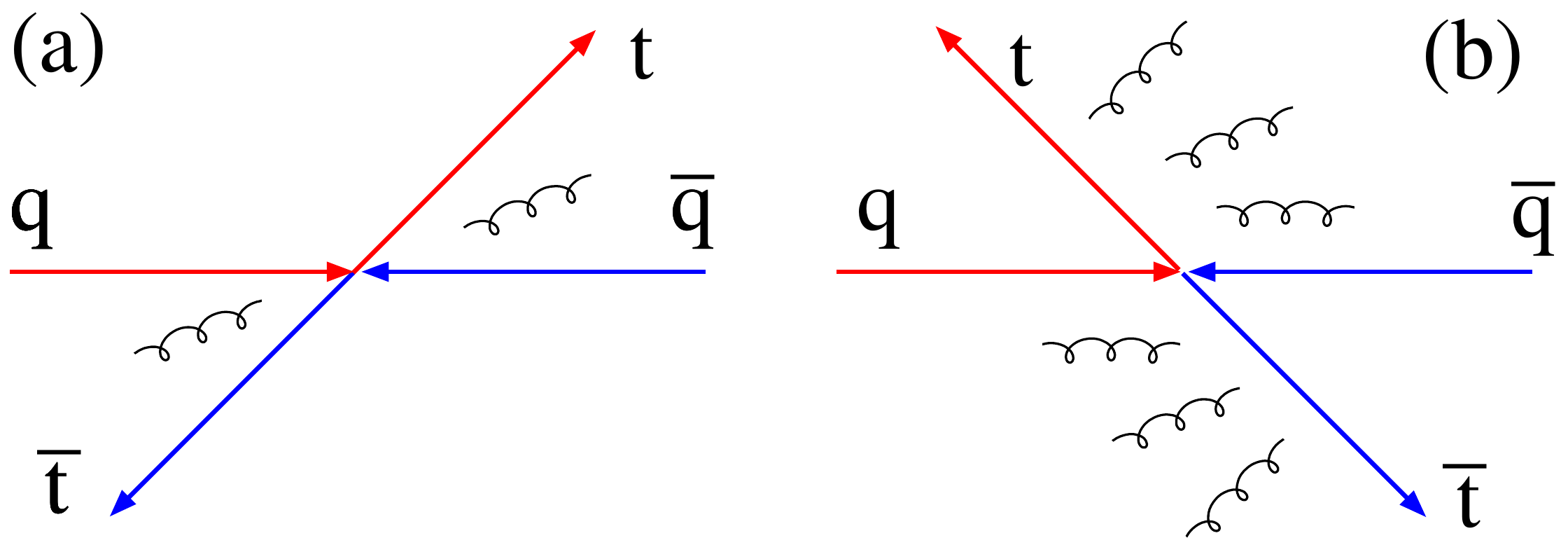}
  \vspace*{-1mm}
  \caption{\label{fig:ttbar_coh}%
    Colour flow and QCD radiation in (a) forward and (b) backward
    $t\bar t$\/ production.}
}

Event generators with coherent parton showers, implemented through 
dipole showering in \Sherpa\ and angular ordering in \Herwig++, 
take account of these effects. (\Pythia\ uses a hybrid between the
two.) A full NLO treatment, included in \MCNLO\ but not in the
stand-alone generators, adds a finite positive virtual
contribution. Nonetheless, as we shall demonstrate in \sec{sec:ps},
even LO shower models can also generate a net inclusive asymmetry
$A_\mathrm{FB}$, if the shower kinematics allow for migration between
positive and negative $\Delta y$\/ regions.

In the following section we examine in more detail the approximations
made in event generators, in comparison to the fixed-order
perturbative treatment.  Then in \sec{sec:incl_gen} we explain in general
terms how they can produce a positive inclusive asymmetry while only
containing the LO production process. In \sec{sec:ps} we present
results from the \Herwig++, \Pythia\ and \Sherpa\ generators for the inclusive
asymmetry and various differential asymmetry distributions. In
\sec{sec:conclude} we summarize our findings and comment on their implications.

\section{Comparison with fixed order}

To establish notation we first consider the lowest-order process,
\beq
q(p_1)+\bar q(p_2)\;\to\;Q(p_3)+\bar Q(p_4)~,
\eeq
for which the leading-order spin-averaged matrix element squared is 
\beq
\overline\sum\,\big|M(q\bar q\to Q\bar Q )\big|^2\;=\;
g^4\,\frac{C_F}N\left(\frac{\bar t^{\,2}+\bar u^2}{\bar s^2}+\frac{2\,m^2}{\bar s}\right)
\eeq
where $m$\/ is the heavy quark mass and
\beq
\bar s\;=\; 2\,p_1\cdot p_2\,,\quad
\bar t\;=\;-2\,p_1\cdot p_3\,,\quad
\bar u\;=\;-2\,p_1\cdot p_4\,.
\eeq
The corresponding differential cross section,
\beq\label{eq:born}
\frac{d\hat\sigma_B}{d\bar t}\;=\;\frac 1{16\pi\,\bar s^2}\,\overline\sum\,
\big|M(q\bar q\to Q\bar Q )\big|^2~,
\eeq
is used for the primary hard subprocess in the event generators.
Clearly, it does not  exhibit any forward--backward asymmetry.   Thus
for an asymmetry to be produced by a leading-order generator, some
parton showering must occur. 

\subsection{One gluon emission}
The leading-order shower contribution is the one-gluon emission process,
\beq
q(p_1)+\bar q(p_2)\;\to\;Q(p_3)+\bar Q(p_4)+g(k)~.
\eeq
For the asymmetry we require the difference between this and the process
\beq
q(p_1)+\bar q(p_2)\;\to\;\bar Q(p_3)+Q(p_4)+g(k)~.
\eeq
The difference between the spin-averaged matrix elements squared
is~\cite{Ellis:1986ef,Kuhn:1998jr,Kuhn:1998kw}
\beqn \label{eq:MA}
{\cal M}_A &\equiv& \overline\sum\,\big|M(q\bar q\to Q\bar Q g)\big|^2 
-\overline\sum\,\big|M(q\bar q\to\bar Q  Q g)\big|^2\nonumber\\[3mm]
&=&
g^6\,\frac{C_F(N^2-4)}{N^2}\,\Biggl[\left(\frac{t_1^2+t_2^2+u_1^2+u_2^2}{s_1s_2}
+\frac{2\,m^2}{s_1}+\frac{2\,m^2}{s_2}\right)\nonumber\\[1mm]
&&\times\ \big(W_{13}+W_{24}-W_{14}-W_{23}\big)-\frac{8\,m^2}{s_1s_2}
\left(\frac{t_1-u_2}{v_2}+\frac{t_2-u_1}{v_1}\right)\Biggr]
\eeqn
where
\beqn
&&
s_1\;=\;(p_1+p_2)^2\,,\quad t_1\;=\;-2\,p_1\cdot p_3\,,\quad
u_1\;=\;-2\,p_1\cdot p_4\,,\quad v_1\;=\;2\,p_3\cdot k\,,\nonumber\\
&&
s_2\;=\;(p_3+p_4)^2\,,\quad t_2\;=\;-2\,p_2\cdot p_4\,,\quad
u_2\;=\;-2\,p_2\cdot p_3\,,\quad v_2\;=\;2\,p_4\cdot k\,,
\eeqn
and $W_{ij}$ is the dipole radiation function
\beq
W_{ij}\;=\;-\left(\frac{p_i}{p_i\cdot k} - \frac{p_j}{p_j\cdot k}\right)^2~.
\eeq
The asymmetries for the processes $gq\to Q\bar Q q$\/ and $g\bar q\to
Q\bar Q \bar q$\/ are obtained from the same expression after crossing.
They are very small and will be neglected in the following.

We see from \eq{eq:MA} that the asymmetry vanishes for $N=2$.  This
must be the case to all orders,
because the fundamental representation of  SU(2) is pseudoreal and
so in that case $q\bar q\to Q\bar Q X$\/ is the same as $q q\to Q Q X$.

In the event generators the gluon radiation is represented as coherent
emission from the external lines of the Born process and so the new colour factor
$(N^2-4)/N$\/ is approximated by $2\,C_F=(N^2-1)/N$.  Thus we expect them to
overestimate the asymmetry at non-zero $p_T\equiv p_{T,Q\bar Q}$ by
around 60\% in lowest order. They also neglect the second term in the
square bracket in  \eq{eq:MA}, which is less singular
at small $k$, and approximate the remaining terms by the Born term
times dipole-like factors.  Thus they effectively treat the asymmetry
in the soft gluon limit.  They treat the gluon radiation more
accurately in the collinear regions, but those regions are not
dominant in the asymmetry.

\subsection{Soft gluon limit}
As explained above, we expect the event generators to reproduce the
soft gluon limit of the asymmetry, apart from the colour factor
mentioned earlier.
In the limit of small $k$\/ we have $s_1,s_2 \to \bar s$\/ etc. and
\eq{eq:MA} takes the simple form
\beqn \label{eq:MAsoft}
{\cal M}_A &=&
g^6\,\frac{(N^2-1)(N^2-4)}{N^3}\,\left(\frac{\bar t^{\,2}+\bar u^2}{\bar s^2}
+\frac{2\,m^2}{\bar s}\right)\nonumber\\[2mm]
&&\times\ \big(W_{13}+W_{24}-W_{14}-W_{23}\big)~.
\eeqn
In the soft gluon limit we can write the differential
cross section for emission of a gluon with energy $\omega$\/ as
\beq
\frac{d^2\hat\sigma}{d\bar t\,d\omega}\;=\;\frac\omega{(4\pi)^4\,\bar s^2}\,
\int\overline\sum\,\big|M(q\bar q\to Q\bar Q g)\big|^2\,d\Omega~,
\eeq
where $\Omega$\/ is the solid angle for soft gluon emission, and we can
use the fact that~\cite{Marchesini:1989yk}
\beq
\int W_{ij}\,d\Omega\;=\;\frac{4\pi}{\omega^2}\,\left[
\frac 1{v_{ij}}\,\log\left(\frac{1+v_{ij}}{1-v_{ij}}\right)
-2\right]\;\equiv\;\frac{4\pi}{\omega^2}\,F_{ij}~,
\eeq
where $v_{ij}$ is the relative velocity of $i$\/ and $j$:
\beq
v_{ij}\;=\;\sqrt{1-\left(\frac{m_i\,m_j}{p_i\cdot p_j}\right)^2}~.
\eeq
Now we define the asymmetry cross section $d\hat\sigma_A\;=\;d\hat\sigma\,-
\,d\hat\sigma\,(3\leftrightarrow4)$:
\beqn\label{eq:sigasy}
\frac{d^2\hat\sigma_A}{d\bar t\,d\omega} &=&   \frac{\as^3}{\bar
  s^2\omega}\,\frac{(N^2-1)(N^2-4)}{N^3}\,\left(\frac{\bar t^{\,2}
+\bar u^2}{\bar s^2}+\frac{2\,m^2}{\bar s}
\right)\big(F_{13}+F_{24}-F_{14}-F_{23}\big)\nonumber\\[2mm]
&=&\frac{2\,\as}{\pi\omega}\,\frac{(N^2-4)}N\,
\big(F_{13}+F_{24}-F_{14}-F_{23}\big)\,\frac{d\hat\sigma_B}{d\bar t}~.
\eeqn
The radiation functions $F_{ij}$  appearing in \eq{eq:sigasy} have
collinear divergences
along the beam directions, which cancel in the full expression.  Regulating
them with a small light quark mass $\mu$, we have in the limit $\mu\to 0$:
\beq
F_{ij}\;\to\;2\log\left(\frac{2\,p_i\cdot p_j}{\mu\,m}\right)-2~,
\eeq
so that
\beq\label{eq:sigasy2}
\frac{d^2\hat\sigma_A}{d\bar t\,d\omega}\;=\;\frac{4\,\as^3}{\bar
  s^2\omega}\,\frac{(N^2-1)(N^2-4)}{N^3}\,\left(\frac{\bar t^{\,2}
+\bar u^2}{\bar s^2}+\frac{2\,m^2}{\bar s}\right) \log\left(\frac{\bar t}{\bar u}\right)~.
\eeq

\FIGURE[!t]{
  \centering
  \includegraphics[scale=0.5]{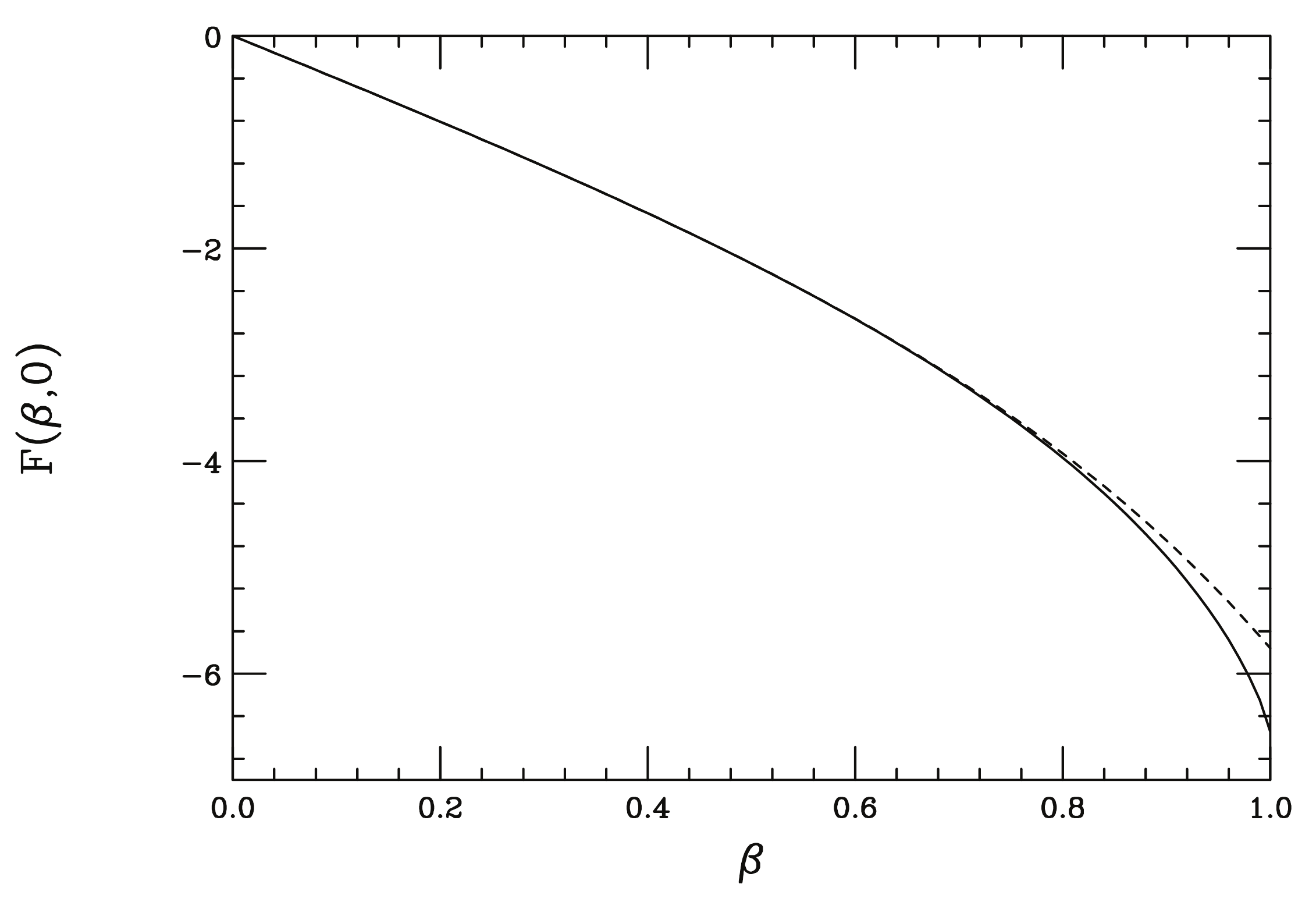}
  \caption{\label{fig:Fbeta}%
    The function $F(\beta,0)$ in \eq{eq:Fbeta}: the full
    expression (solid) and the first four terms of its power series
    expansion (dashed).}
}

In lowest order the $Q\bar Q$\/ pair recoils against the emitted gluon, so
the transverse momentum of the pair is $p_T=-k_T$ where $dk_T/k_T =
d\omega/\omega$.  We can also express \eq{eq:sigasy2} in terms of
the $q\bar q\to Q\bar Q$\/ scattering angle $\hat\theta$\/ since
\beq\label{eq:bart}
\bar t\;=\;-\bar s-\bar u\;=\;-\frac 12\,\bar s\,(1-\beta\cos\hat\theta)~,
\eeq
where $\beta=\sqrt{1-4\,m^2/\bar s}$\/ is the heavy quark c.m.~velocity
in the Born process.  Thus
\beq\label{eq:sigasy3}
\frac{d^2\hat\sigma_A}{dp_T\,d\cos\hat\theta}\;=\;\frac{\as^3\,\beta}{\bar
  s\,p_T}\,\frac{(N^2-1)(N^2-4)}{N^3}\,\left(2-\beta^2\sin^2\hat\theta\right)
\log\left(\frac{1-\beta\cos\hat\theta}{1+\beta\cos\hat\theta}\right)~.
\eeq
Finally, we can integrate this expression to find the recoil
distribution of the asymmetry in the soft limit.  It is convenient to
normalize this to the $q\bar q\to Q\bar Q$\/ Born cross
section,\footnote{At the Tevatron, the gluon fusion contribution to
  the $t\bar t$\/ production cross section is small, less than 10\%
  (at LO, we find $6.7$\% using CTEQ6L1 PDFs \cite{Pumplin:2002vw} and
  scales evaluated via the top quark transverse mass).}
obtained by integrating \eq{eq:born}:
\beq
\hat\sigma_B\;=\;\frac{\pi\as^2}{24\,m^2}\,
\frac{(N^2-1)}{N^2}\,\beta\,(1-\beta^2)\,(3-\beta^2)~.
\eeq
Then we can write in general
\beq\label{eq:sigasy4}
\frac{p_T}{\hat\sigma_B}\,\frac{d\hat\sigma_A}{dp_T}\;=\;\frac{\as}{\pi}\,\frac{(N^2-4)}{N}\,F(\beta,p_T)~,
\eeq
where the coupling- and colour-stripped asymmetry function
$F(\beta,p_T)$ is given in the soft limit,
$p_T\ll p^\mathrm{max}_T=\beta^2\sqrt{\bar s}/2$, by
\beqn\label{eq:Fbeta}
F(\beta,0)&=&\frac{6}{(3-\beta^2)}\,\int_0^1\left(2-\beta^2 +\beta^2 z^2\right)
\log\left(\frac{1-\beta z}{1+\beta z}\right)\,dz\nonumber\\[2mm]
&=&\frac 2{\beta\,(3-\beta^2)}\,\left[(1-\beta)\,(7+\beta-2\,\beta^2)\log\left(\frac{1+\beta}{1-\beta}\right)
-2\,(7-3\,\beta^2)\log(1+\beta)-\beta^2\right]\nonumber\\[2mm]
&=& -4\,\beta -\beta^3-\frac{22}{45}\,\beta^5 -\frac{103}{378}\,\beta^7 +{\cal O}(\beta^9)~.
\eeqn
Thus we expect the asymmetry in the event generators to become more
negative with increasing top pair invariant mass $\sqrt{\bar s}$,
growing linearly with c.m.~velocity $\beta$\/ near threshold.   The
function $F(\beta,0)$ is shown in \fig{fig:Fbeta}.  It tends to an
asymptotic value of $-8\log 2-1 = -6.545$ as $\beta\to 1$.  We note
that the four-term power series expansion \eq{eq:Fbeta} gives a
good approximation over a wide range of $\beta$.

\subsection{Beyond the soft approximation}\label{sec:nonsoft}
The coupling- and colour-stripped asymmetry function $F(\beta,p_T)$
remains negative away from the soft region but decreases in magnitude
as $p_T$ increases, vanishing at the phase-space boundary, as shown in
\fig{fig:FbetapT}.  Here, as before, $\beta$\/ is defined as
$\sqrt{1-4\,m^2/s_1}$ , i.e.\ in terms of the overall centre-of-mass
energy squared $s_1=\bar s$, rather than the invariant mass squared
of the $Q\bar Q$\/ pair, $s_2$.

\FIGURE[!t]{
  \centering
  \includegraphics[scale=0.5]{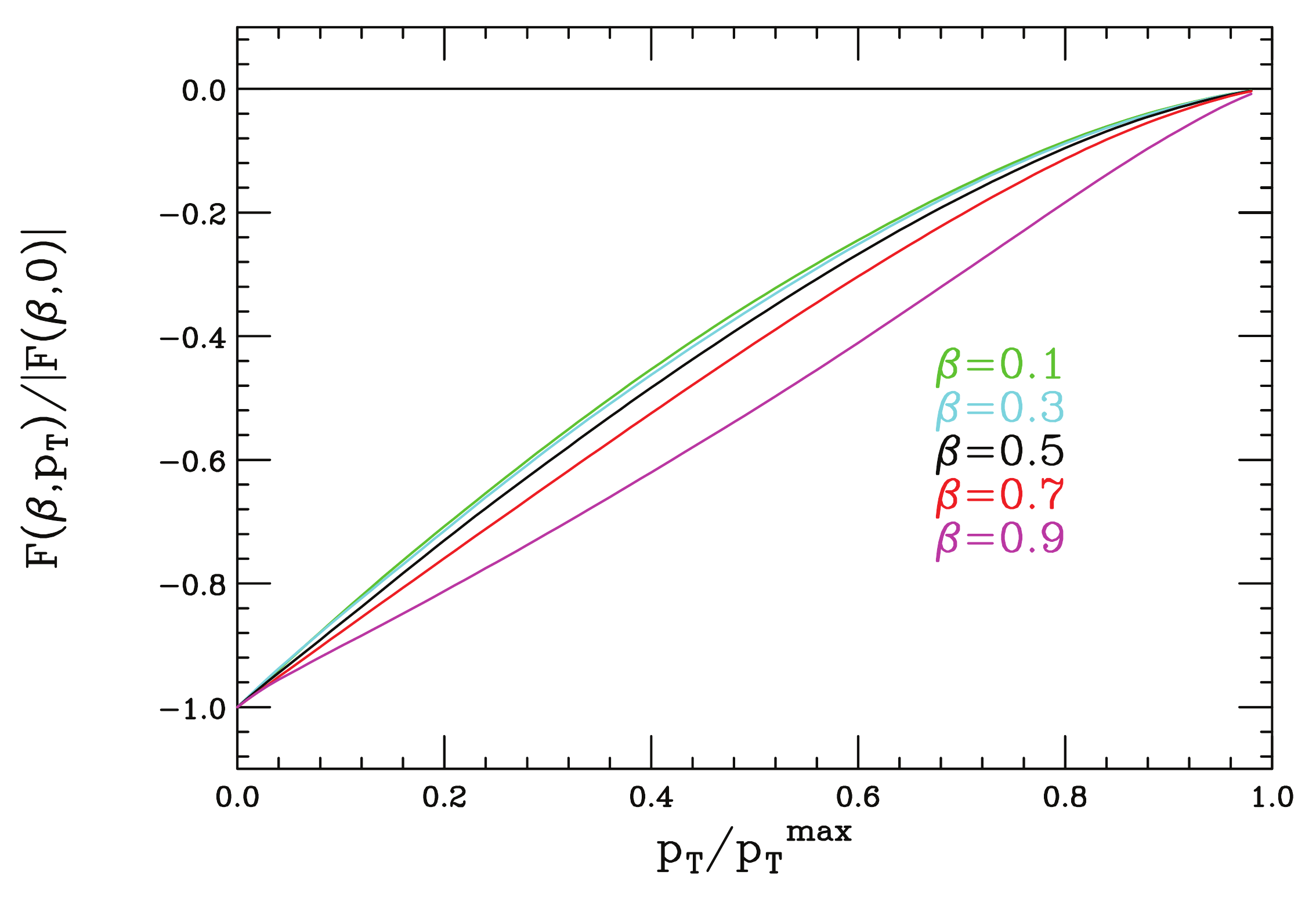}
  \caption{\label{fig:FbetapT}%
    The function $F(\beta,p_T)$, relative to its value at
    $p_T=0$. Note that $p^\mathrm{max}_T=\beta^2\sqrt{\bar s}/2$.}
}

To go from this function to the $p_T$-dependence of the asymmetry,
$A_{\rm FB}(p_T)$, as defined in \eq{eq:AFBdef}, we have to include the
strong coupling and colour factors and normalize to the differential
cross section at the same $p_T$.  A full leading-order calculation of
the asymmetry using \textsc{MCFM}~\cite{Campbell:2010ff} (computing
$t\bar t$\/ production at NLO) yields the results shown in
\fig{fig:LO_pt_asy}.  For this calculation, the next-to-leading order
parton distribution functions of Ref.~\cite{Martin:2009iq} were used,
with scale equal to the top mass ($m_t=172.5$~GeV), but the results
for the asymmetry are rather insensitive to these choices.

The real-emission asymmetry cross section in the numerator of
\eq{eq:AFBdef} diverges as $p_T\to 0$, due to the soft
divergence discussed in the previous section.  However, the
denominator, shown on the left in \fig{fig:LO_pt_asy}, 
diverges faster, as it has initial-state collinear
singularities that cancel in the numerator, so that $A_{\rm FB}(p_T)$ 
is driven towards zero as $p_T\to 0$.

In the fully inclusive asymmetry, the divergence of the real emission
contribution is cancelled by the singular virtual correction at
$p_T=0$, and we have
\beq
\frac 1{\hat\sigma_B}\,
\int_0^m\frac{d\hat\sigma_A}{dp_T}\,dp_T\;=\;
\frac{\as}{\pi}\,\frac{(N^2-4)}{N}\,f(\beta)~,
\eeq
where $f(\beta)$ is finite.  If we limit the
integration to $p_T<q_T\ll m$, we therefore find
\beqn\label{eq:sigAqt}
\frac{\hat\sigma_A(p_T<q_T)}{\hat\sigma_B} &=&
\frac 1{\hat\sigma_B}\,\left(
\int_0^m\frac{d\hat\sigma_A}{dp_T}\,dp_T\;-\;
\int_{q_T}^m\frac{d\hat\sigma_{\rm A}}{dp_T}\,dp_T \right)\nonumber\\[2mm]
&\sim&\frac{\as}{\pi}\,\frac{(N^2-4)}{N}\,
\Bigg[F(\beta,0)\log\left(\frac{q_T}m\right)\,+\,G(\beta,q_T/m)\Bigg]
\eeqn
where $F(\beta,0)$ is given by \eq{eq:Fbeta} and $G(\beta,q_T/m)$ is
regular at $q_T=0$.
Since $F(\beta,0)\sim-4\,\beta$\/ is negative, a cut on $p_T<q_T$ adds a
positive contribution to the inclusive asymmetry, which grows
logarithmically as $q_T$ is reduced.   This effect can be quite
significant.  For example, in top pair
production at a pair invariant mass of 450~GeV ($m_t=172$~GeV,
$\beta=0.645$), a cut on $p_T<20$~GeV gives a contribution of
$3.33\,\as\sim35$\% from the logarithmic term.

\FIGURE[!t]{
  \centering\centerline{
  \includegraphics[scale=0.57]{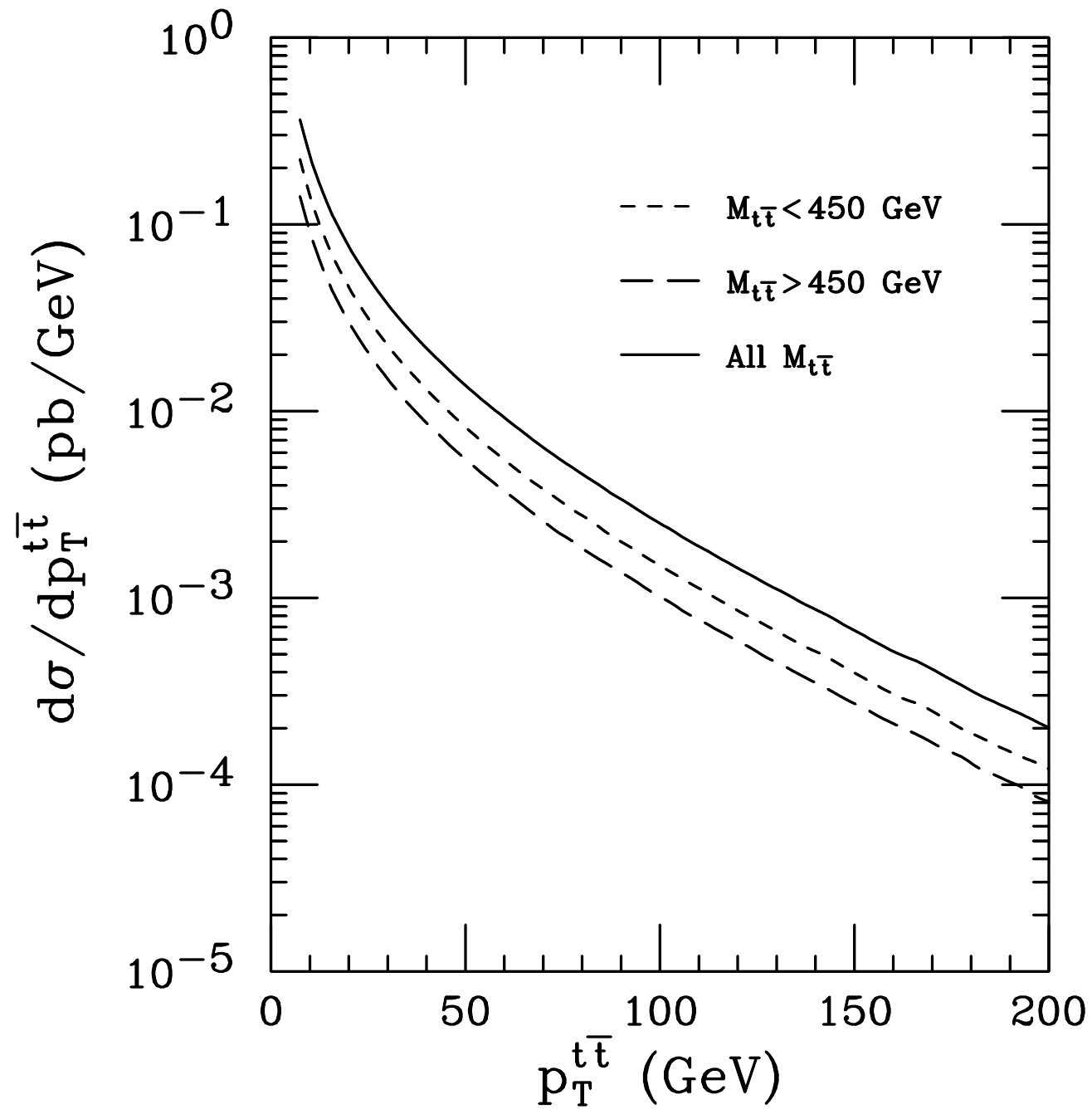}
  \includegraphics[scale=0.57]{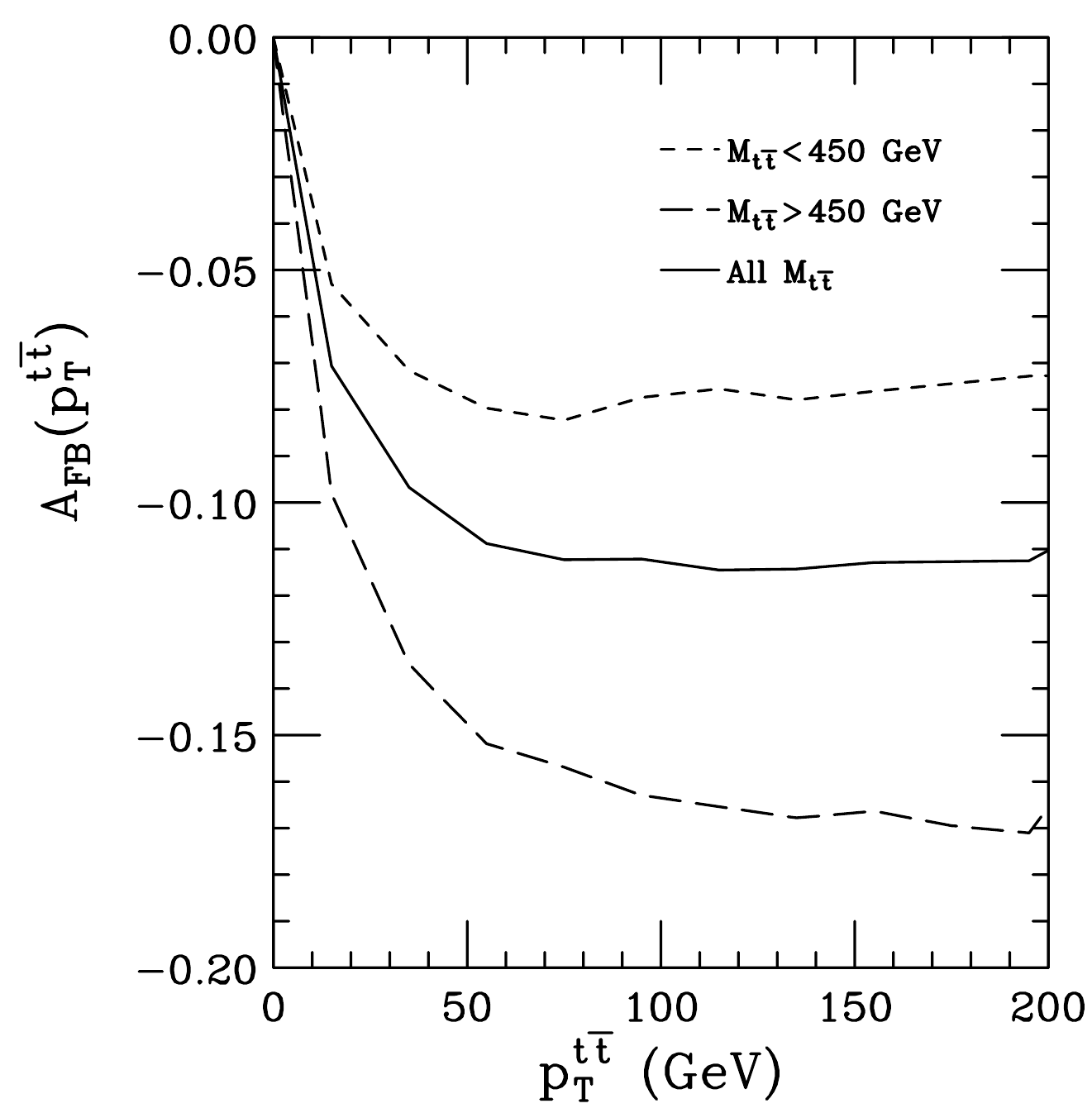}}
  \vspace*{-17pt}
  \caption{\label{fig:LO_pt_asy}%
    Leading-order QCD predictions for the transverse-momentum
    distribution of the top quark pair, $p_T^{t\bar t}$ (left),
    and for the forward--backward asymmetry as a function of
    $p_T^{t\bar t}$ (right).}
}

In a combined NLO plus parton shower treatment such as \MCNLO, the
positive singular virtual contribution at $p_T=0$ is spread out over a
finite region of $p_T$, the so-called Sudakov region.  This leads to a
cross-over in the asymmetry from positive values at low $p_T$
to negative values at higher $p_T$, as depicted in
\fig{fig:D0_pt_asy}.  In \MCNLO, the finite part of the virtual
contribution, absent from the event generators, is also included and
can affect the position of the cross-over.

Monte Carlo event generators with QCD coherence will not have the
correct form for the function $G$\/ in \eq{eq:sigAqt}, due to the
inexact treatment of hard, non-collinear emissions, but they
should reproduce the logarithmic term, apart from the overestimate of
the colour factor by 60\% mentioned earlier.  They will also display
the spreading of the positive asymmetry over the Sudakov region at low
$p_T$.

\section{Generation of an inclusive asymmetry}\label{sec:incl_gen}

Despite the coherence effect elaborated upon in the previous section, the
naive expectation is that the asymmetry should still sum
to zero when integrated over all of phase space.\footnote{Or, if using
  \MCNLO\ or \POWHEG~\cite{Nason:2004rx,Frixione:2007nw}, that
 the integrated forward--backward asymmetry should sum to its NLO
 inclusive value.} This expectation is based on the simple fact that
showers are unitary (meaning real-radiation corrections cancel exactly
against virtual-Sudakov ones), so even though they can move things
around in phase space, they do not generate any corrections to total
cross sections. At the most inclusive level, this is reflected in the
fact that the total integrated $t\bar{t}$\/ cross section is the same
before and after showering.

However, the asymmetry is defined in terms of two separate cross
sections, one computed for $\Delta y > 0$ and the other for $\Delta y
< 0$. If the shower kinematics allow any migration between these two
regions, then unitarity no longer guarantees complete cancellation in
each of the regions separately, leading to the possible generation of
a net inclusive asymmetry. Formally, we can write the cross
section difference that generates the integrated asymmetry as
\begin{eqnarray}
\Delta\sigma_{+-} & = &
\int d\sigma^{\mathrm{LO}}\big\rfloor_{\Delta y > 0} \ \big[\,\Delta_+ + (1-\Delta_+)(P_{++} - P_{+-})\,\big] \nonumber\\[1mm]
& &\quad -\int d\sigma^{\mathrm{LO}}\big\rfloor_{\Delta y < 0} \ \big[\,\Delta_- + (1-\Delta_-)(P_{--} - P_{-+})\,\big]~, \label{eq:incl}
\end{eqnarray}
where the first line represents events that start (at the
matrix-element level, before showering) 
with a positive value of $\Delta y$\/ and the second line
represents events that start with a negative one. The terms in
parentheses represent the action of the parton shower. The probability
for no branchings to occur is represented by the Sudakov factor,
$\Delta$, with subscript $\pm$ reflecting that 
the probability to radiate can be
different between an event with positive $\Delta y$\/ and one with
negative $\Delta y$.  Indeed,  as shown in the preceding section,
events with  positive $\Delta y$\/ have less phase space
for emission and so are less likely to radiate.  Therefore, in general, we have
\begin{equation}\label{eq:deltapm}
\Delta_+\;>\;\Delta_- ~.
\end{equation} 
This, however, is not by itself enough to generate an inclusive
asymmetry. The second terms in the square brackets in \eq{eq:incl}
represent those events that do experience one or more branchings. For
these events, the final top momenta, and hence possibly their final
rapidity difference, will depend on whether and how the top momenta
are modified by the branchings. In the present
context, we do not care about the details of how this occurs, 
merely about whether it is at all possible for an
event with positive $\Delta y$\/ at the Born level
to migrate to negative $\Delta y$\/ after showering, and vice versa. 
This is represented by the
probabilities $P_{+-}$ and $P_{-+}$ in \eq{eq:incl}. If
the shower model preserves the rapidity ordering of the tops, then 
\begin{equation}
P_{++}\;=\;P_{--} = 1 ~~~~~~\mbox{and}~~~~~~ P_{+-}\;=\;P_{-+}\;=\;0~,
\end{equation}
and so the integrated asymmetry remains zero, despite the two Sudakov
factors being different. If, on the other hand, the shower model
sometimes changes the relative rapidity ordering of the tops,
for instance as a consequence of longitudinal recoil
effects (as will be studied in more detail in the next section), 
then a total inclusive asymmetry can be generated.
In the context of unitarity, this can be interpreted as 
due to the fact that unitarity involves
an integral over the entire phase space, and hence the exact
cancellation that occurs in the total inclusive cross section is here
broken by the splitting-up of the real-radiation phase space into two regions
that enter with different signs in the asymmetry. 

From unitarity of the shower, we have
\begin{equation}
P_{++}\;=\;1-P_{+-} ~~~~~~\mbox{and}~~~~~~ P_{--}\;=\;1-P_{-+}~,
\end{equation}
so that \eq{eq:incl} can be written as
\begin{equation}
\Delta\sigma_{+-}\;=\;
 -\;2\int d\sigma^{\mathrm{LO}}\big\rfloor_{\Delta y > 0} \ (1-\Delta_+) P_{+-}
 \;
 +\;2\int d\sigma^{\mathrm{LO}}\big\rfloor_{\Delta y < 0} \ (1-\Delta_-) P_{-+}~, 
\label{eq:incl2}
\end{equation}
where we have used
\begin{equation}
  \int d\sigma^{\mathrm{LO}}\big\rfloor_{\Delta y > 0}
  \;=\;\int d\sigma^{\mathrm{LO}}\big\rfloor_{\Delta y < 0}~.
\end{equation}
Because $1>\Delta_+>\Delta_-$, we expect the second term on the right-hand side
of \eq{eq:incl2} to dominate, giving a positive inclusive
asymmetry, unless there is a compensating excess of $ P_{+-}$  over
$P_{-+}$.  However, on rather general grounds one would not expect
such an excess, because there is less radiation when $\Delta y > 0$
and hence a smaller probability of recoil effects changing the  
sign of $\Delta y$.  Indeed we shall see below that the treatment
of recoils in shower generators normally leads to $P_{-+}>P_{+-}$,
enhancing the positive inclusive asymmetry due to the unequal
Sudakov factors.

Considering \eq{eq:incl2} from the viewpoint of perturbation theory,
we observe that the factors of $(1-\Delta_\pm)$ in the integrands are
${\cal O}(\as^1)$, while $P_{\pm\mp}$ are ${\cal O}(\as^0)$, being the
{\it conditional}\/ probabilities that gluon emission will switch the sign
of $\Delta y$, given that at least one emission has occurred. Thus the
recoil effect in showering generates an approximate inclusive asymmetry
that starts at ${\cal  O}(\as)$, like the full perturbative calculation.
The factors of $(1-\Delta_\pm)$ provide information about the virtual
contribution and the probabilities $P_{\pm\mp}$ specify what fraction
remains after real-virtual cancellation.  Since these probabilities depend
on the strategy for treating recoils in the shower, getting the best
agreement with the full asymmetry at ${\cal O}(\as)$ could be
a good way to optimize this strategy.

\section{Comparison between parton-shower models \label{sec:ps}}

In this section, we study the asymmetries produced by the following
general-purpose event generators:\,\footnote{Specifically, we use
  \Herwig++ 2.5.2, \Pythia~6.426, \Pythia~8.162, and \Sherpa\ 1.4.0.}
\Herwig++~\cite{Bahr:2008pv} (using angular-ordered parton
showers~\cite{Gieseke:2003rz}),
 \Pythia~6~\cite{Sjostrand:2006za} (using both its $Q^2$- and
$p_\perp$-ordered parton-shower
  models~\cite{Sjostrand:1985xi,Sjostrand:2004ef}, represented by
  tunes D6T and Perugia 0, respectively),
\Pythia~8~\cite{Sjostrand:2007gs} (using
$p_\perp$-ordered parton showers~\cite{Corke:2010zj}), 
and \Sherpa~\cite{Gleisberg:2008ta} (using $p_\perp$-ordered dipole
showers~\cite{Schumann:2007mg}). Of these, \Herwig++\ and \Sherpa\
have QCD coherence built in and \Pythia~6 has options
with varying amounts of coherence, while the first ISR (initial-state
radiation) emission is not subjected to coherence constraints in this
version of \Pythia~8. For both \Pythia~6 and \Sherpa, we include some
additional illustrations of specific shower model variations.

A custom-made RIVET~\cite{Buckley:2010ar} analysis was used to process
the events of all generators, ensuring uniformity of the
analysis. Between 1 and 4 million events (at least) were generated for
each model. All the generators include the leading-order 
$q\bar{q}\to t\bar{t}$\/ and $gg\to t\bar{t}$\/ production processes,
which are showered with default settings,\footnote{The choice of PDF
  set only gives small effects ($\lesssim10$\%) on the asymmetry,
  mostly via the relative fraction of gluon-initiated vs.\
  quark-initiated $t\bar{t}$\/ production. For completeness, \Herwig++
  uses the MRSTMCal PDF set (i.e.\ the LO fit from the MRST2002 family)
  \cite{Martin:2002dr}, \Pythia~6 with Perugia 0 uses CTEQ5L
  \cite{Lai:1999wy}, and \Pythia~6 with D6T, \Pythia~8, and \Sherpa\
  all use CTEQ6L1 PDFs \cite{Pumplin:2002vw}. There is also a slight
  dependence on the choice of renormalization scale, see~\app{app:addmat}.}
unless otherwise specified. Note that we do not study the effects of
matrix-element plus parton-shower matching in this paper.

\TABLE[!t]{
\centering\small
\begin{tabular}{lllclcclcc}\hline\hline
  \multicolumn{1}{l}{~Model} &
  \multicolumn{1}{l}{\rule[-3.5mm]{0mm}{9mm}~Version} &
  \multicolumn{1}{l}{\hphantom{.}} &
  \multicolumn{1}{c}{Inclusive} & \multicolumn{1}{l}{\hphantom{.}} &
  \multicolumn{1}{c}{$m_{t\bar t}/\mathrm{GeV}$} &
  \multicolumn{1}{c}{} & \multicolumn{1}{l}{\hphantom{.}} &
  \multicolumn{1}{c}{$p_{T,t\bar t}/\mathrm{GeV}$} &
  \multicolumn{1}{c}{}\\[-1mm]
  &\multicolumn{1}{c}{\rule[-3.0mm]{0mm}{8mm}~~~~~~~[tune]} &&&&
  \multicolumn{1}{c}{$<450$} & \multicolumn{1}{c}{$>450$} &&
  \multicolumn{1}{c}{$<50$} & \multicolumn{1}{c}{$>50$}\\\hline
  \rule[0mm]{0mm}{5mm}\Herwig++ &
              2.5.2\, [def]  && $3.9$  && $2.7$ &$6.0$&&$5.8$&$-14.3$\\[7pt]
  \Pythia~6 & 6.426   [def]  && $-0.1$ && $-0.8$&$1.2$&&$2.5$&$-42.5$\\[0mm]
  \Pythia~6 & 6.426   [D6T]  && $-0.2$ && $-1.1$&$1.2$&&$3.2$&$-43.4$\\[0mm]
  \Pythia~6 & 6.426   [P0]   && $0.8$  && $0.7$ &$1.1$&&$1.8$&$-8.6$\\[0mm]
  \Pythia~8 & 8.163   [def]  && $2.5$  && $2.4$ &$2.8$&&$2.4$&$4.8$\\[7pt]
  \Sherpa   & 1.4.0\, [def]  && $5.5$  && $3.5$ &$9.2$&&$8.7$&$-15.4$\\[0mm]
  \Sherpa   & 1.3.1\, [def]  && $6.3$  && $3.3$&$12.1$&&$9.6$&$-15.8$\\[1mm]\hline
  \rule[0mm]{0mm}{5mm}QCD &
              LO             && $6.0$  && $4.1$ &$9.3$&&$7.0$&$-11.1$\\[1mm]\hline\hline
\end{tabular}
\caption{\label{tab:totasym}%
  The forward--backward asymmetry $A^{\mathrm{(cut)}}_\mathrm{FB}$ (in \%)
  in each shower model, and to leading non-trivial order.
  $A^{\mathrm{(cut)}}_\mathrm{FB}$ is defined in \eq{eq:afbcut}. The
  \Pythia\ tunes are discussed in \sec{sec:pythia.tune}, and the
  shower model used in \Sherpa\ is the \Css\ one. The brackets after
  the generator version number denote the parameter set (tune) used, 
  with [def] for default parameters. The fixed-order predictions are
  from \textsc{MCFM}~\protect{\cite{Campbell:2010ff}} used to compute
  $t\bar t$\/ production at NLO.}
}

\subsection{Inclusive asymmetry}\label{sec:incl_comp}

The inclusive asymmetry and the asymmetries with an invariant mass or
transverse-momentum cut,
\beq\label{eq:afbcut}
A^{\mathrm{(cut)}}_\mathrm{FB}\;=\;
\frac{\;\sigma^{\mathrm{(cut)}}\big\rfloor_{\Delta y>0}\,-\,
  \sigma^{\mathrm{(cut)}}\big\rfloor_{\Delta y<0}\;}
     {\sigma^{\mathrm{(cut)}}}~,
\eeq
produced by each model are given in \tab{tab:totasym}. Of these,
\Sherpa's \Css\ produces the largest inclusive asymmetry. We
interpret this as a consequence of its initial--final dipole
kinematics~\cite{Schumann:2007mg,Hoeche:2009xc,Carli:2010cg};
part of the longitudinal momentum of the first emitted gluon has to
come from the recoiling top quark, changing its rapidity and allowing
$\Delta y$\/ to change sign.  Later on, we will illustrate and discuss
recoil effects in somewhat more detail, in a small \Css\ case study,
see~\sec{sec:css.vary} and \app{app:addmat}.

In \Herwig++, coherence is implemented by angular-ordered parton
branching rather than dipole showering.  Parton showers associated
with each incoming or outgoing hard parton are generated independently in
angular regions defined by the colour structure of the hard
subprocess. The showers are then combined according to a kinematic
reconstruction algorithm~\cite{Bahr:2008pv} that again reflects the
colour structure of the subprocess.  In the case of  $q\bar q\to t\bar
t$, there are two initial--final colour-connected systems, $qt$\/ and
$\bar q\bar t$, as illustrated in \fig{fig:ttbar_coh}.  The resulting
treatment of recoils is similar to that in dipole showering of these systems,
with a particular prescription for sharing recoil momentum within
each.  The separate recoils of the top quark and antiquark again
imply that $\Delta y$\/ may change sign.

In contrast, the approach to initial--final dipoles in \Pythia, for
both $p_\perp$- and $Q$-ordered
showers~\cite{Sjostrand:1985xi,Sjostrand:2004ef}, 
is to use the other incoming parton for momentum
conservation, rather than the recoiling top. The 
relative rapidity ordering of the top and the antitop is normally preserved by
this strategy, resulting in very little net asymmetry being generated.

Overall, the asymmetries in \Herwig++\ and \Sherpa\ are comparable
to the LO perturbative results\,\footnote{NLO calculations of the
  forward--backward asymmetry in $t\bar t+1$-jet production
  \cite{Dittmaier:2008uj,Alioli:2011as} give values some 6\%
  above the LO result in \tab{tab:totasym}, e.g.\ in
  Ref.~\cite{Dittmaier:2008uj} a value of $-3.05$\% is reported for
  $p^\mathrm{jet}_T>50$~GeV.} shown in \tab{tab:totasym},
suggesting that their similar recoil strategies are not far from
optimal.

\subsection{Asymmetry as a function of top quark observables}
\label{sec:excl_comp}

We now show differential spectra $d\sigma/dO$\/ for four
key observables and their related forward--backward asymmetry
distributions $A_\mathrm{FB}(O)$, as defined in \eq{eq:AFBdef}.
The observables presented here are the azimuthal angle $\Delta\phi$\/
between the transverse momenta of the top and antitop quarks,
the $|\Delta y|$ distribution itself, and the transverse-momentum
and invariant mass distributions of the $t\bar t$\/ pair. We take a subset
of the shower versions listed in \tab{tab:totasym} (neglecting the
\Pythia~6 default and \Sherpa\ 1.3.1 versions) and compare their
predictions with each other.

\FIGURE[!t]{
  \centering
  \centerline{\hspace*{-7mm}
    \includegraphics[width=0.492\columnwidth]{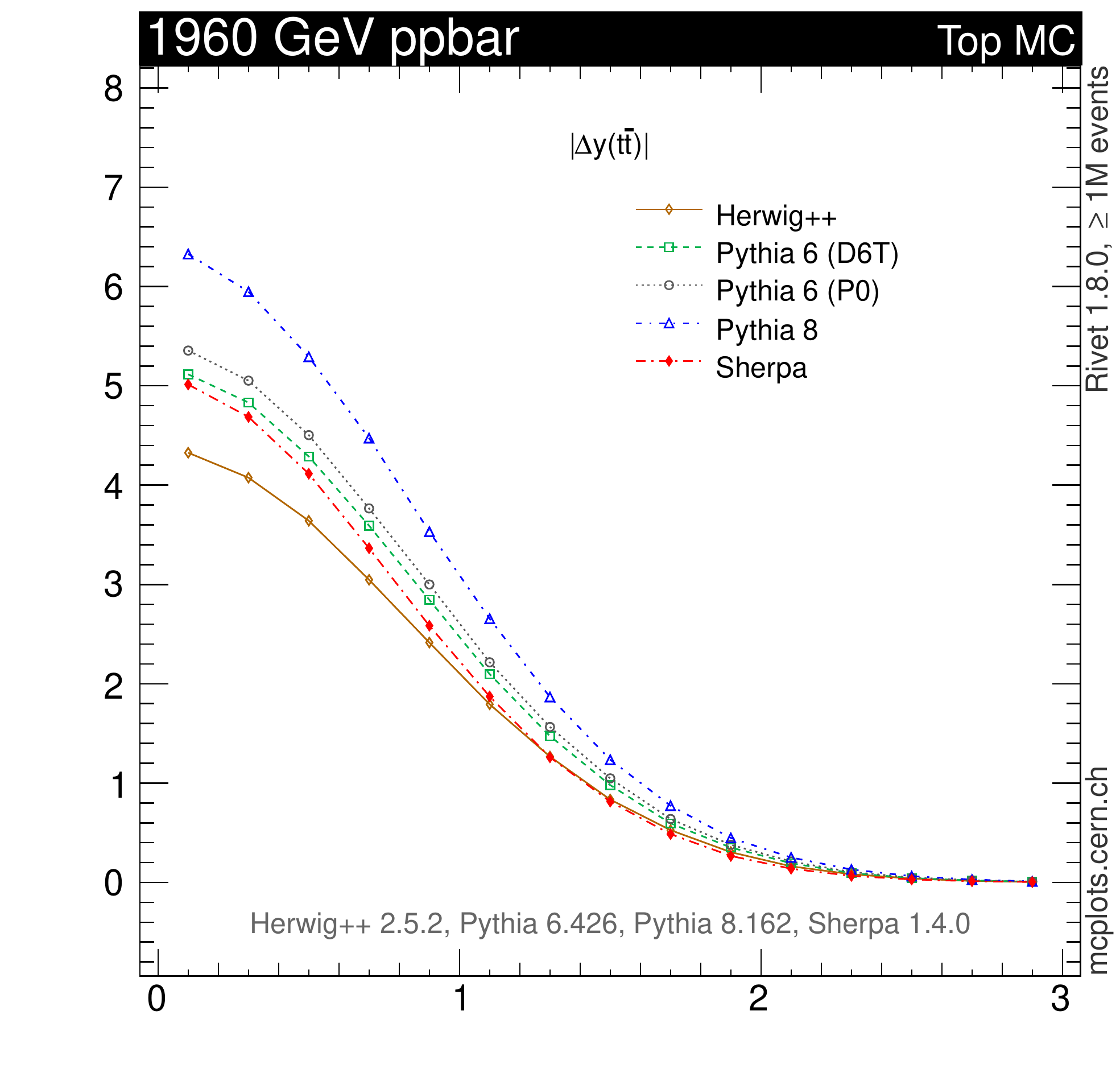}
    \includegraphics[width=0.492\columnwidth]{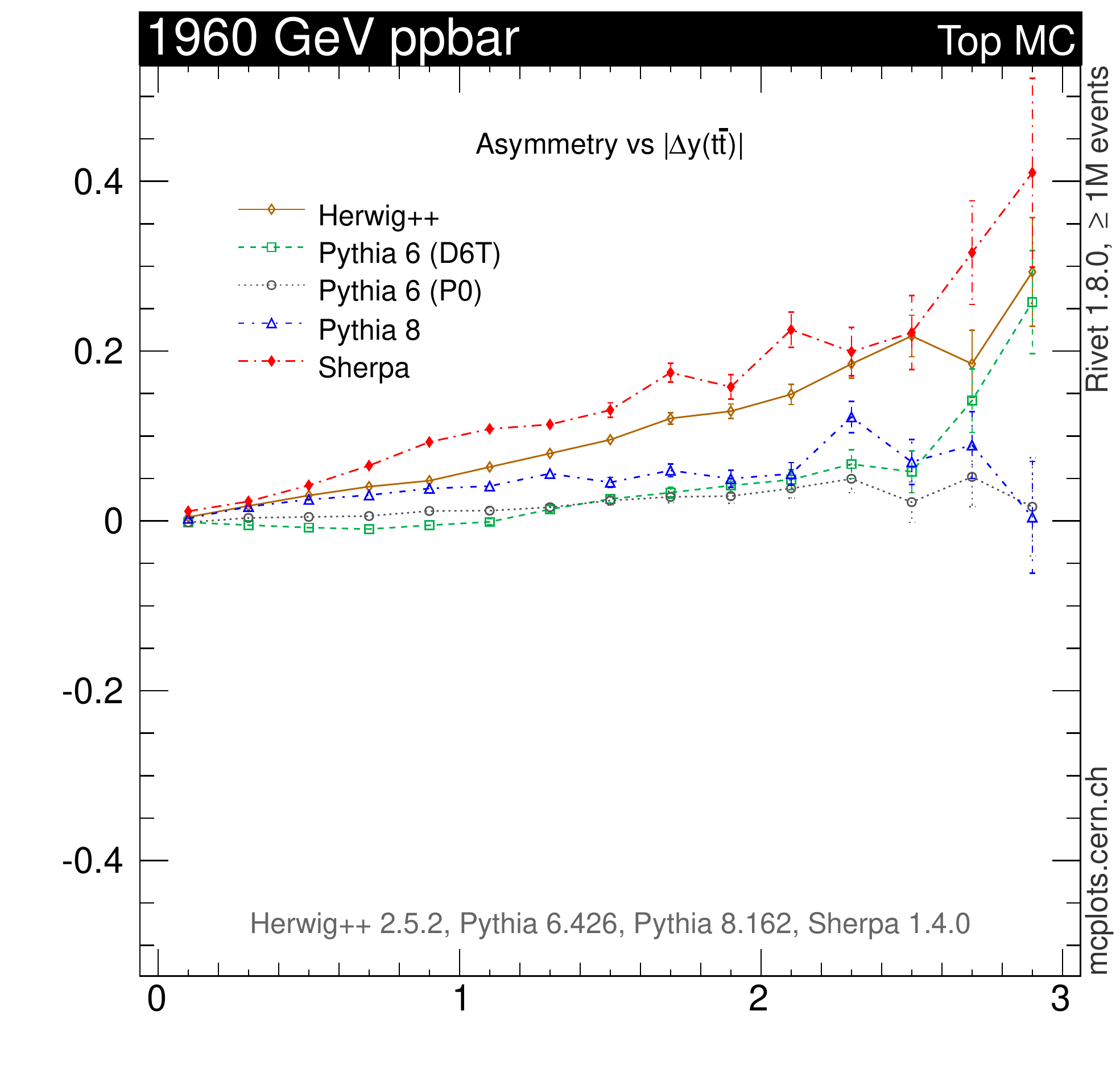}}
  \centerline{\hspace*{-7mm}
    \includegraphics[width=0.492\columnwidth]{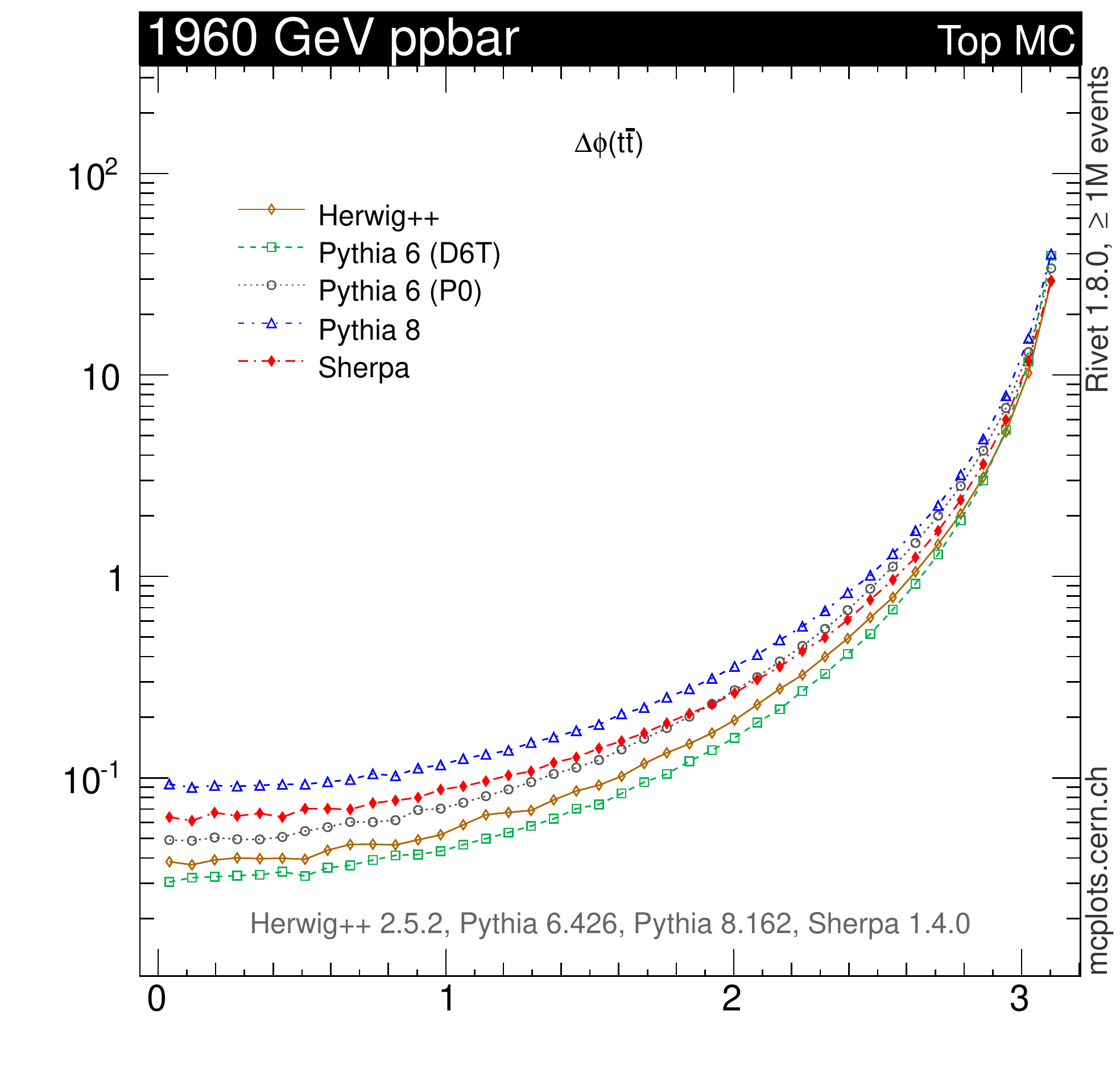}
    \includegraphics[width=0.492\columnwidth]{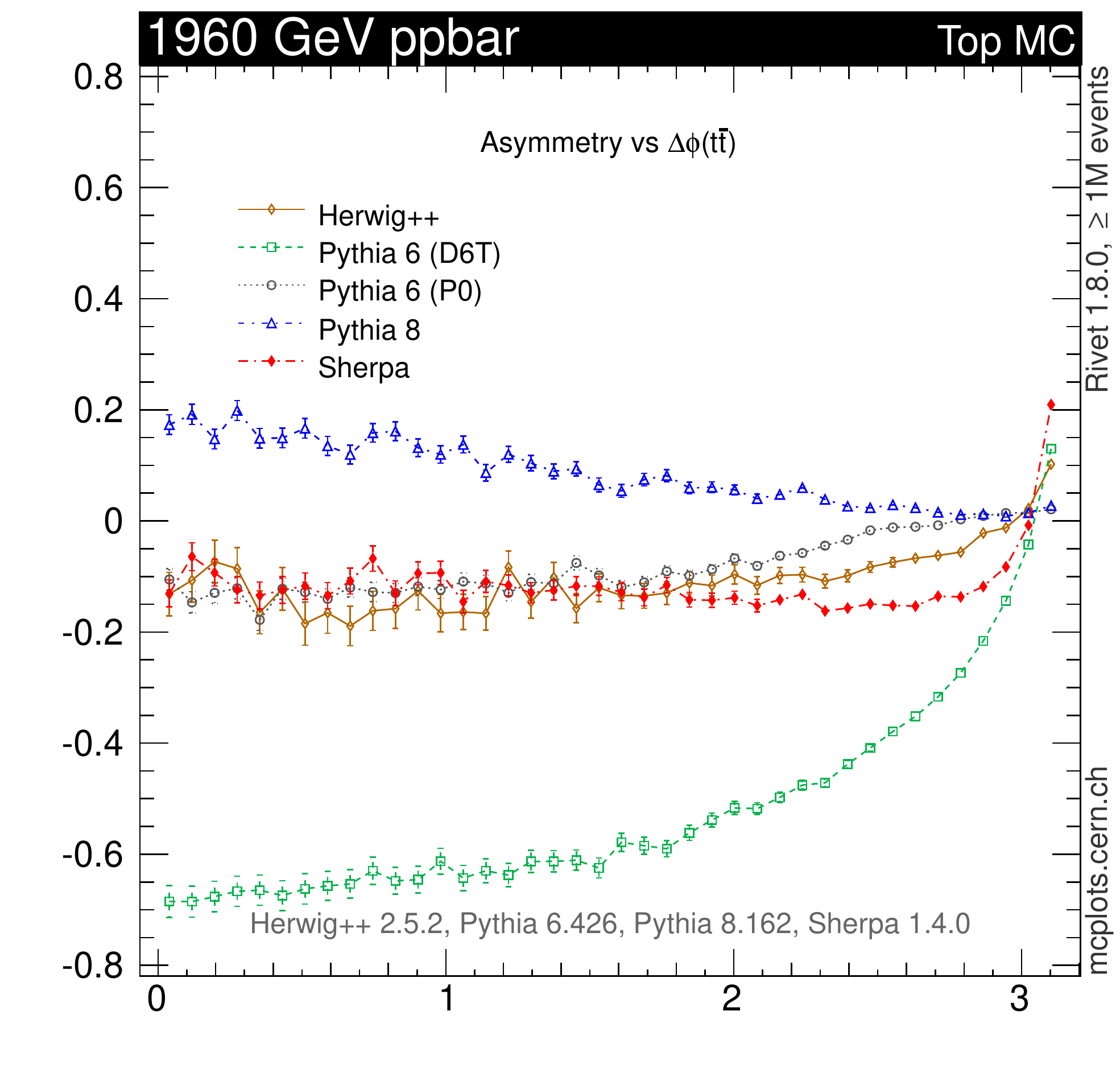}}
  \vspace*{-25pt}
  \caption{\label{fig:tasy.dphitt}%
    Differential cross section (in pb) and top quark forward--backward
    asymmetry $A_\mathrm{FB}$ as a function of the modulus of the
    rapidity difference $\Delta y$\/ (upper row) and the azimuthal
    angle $\Delta\phi$\/ (lower row) between the top and antitop
    quarks. Predictions are shown as obtained from various Monte Carlo
    event generators; errors are statistical only.}
}

In \fig{fig:tasy.dphitt} we show the $|\Delta y|$ and $\Delta\phi$\/
distributions. The $d\sigma/d|\Delta y|$ predictions are very similar
in shape; they differ in normalization because of the spread of the
total inclusive cross sections evaluated at LO and different scales in
the various event generators. The asymmetry rises for larger absolute
rapidity differences of the top quark rapidities. The large $|\Delta y|$
configurations emerge more easily in scatterings with small angle to
the beam (or the $q\bar q$\/ axis). This produces a positive asymmetry
since the associated initial--final $qt$\/ and $\bar q\bar t$\/
dipoles tend to emit softer gluons, with higher rate, in forward
direction.  The mechanism is also explained more fully in
\sec{sec:css.vary}. All \Pythia\ predictions increase rather mildly,
while those of \Herwig++ and \Sherpa\ show a steeper (approximately
linear) slope, which is slightly flatter, but qualitatively comparable
with the recent \textsc{MCFM} results given for the acceptance
corrected case, see Ref.~\cite{Campbell:2012uf}.

The $\Delta\phi$\/ variable is a typical example of an observable
separating the hard-emission domain from the Sudakov region, here
located around large $\Delta\phi\approx\pi$. We depict the
$d\sigma/d\Delta\phi$\/ spectra in \fig{fig:tasy.dphitt} reflecting
the different levels of hardness produced by the different generators
with \Pythia~8 giving the highest levels. Because of the strong
correlation with the $p_{T,t\bar t}$ observable, $\Delta\phi$\/
displays a qualitatively similar behaviour of the related asymmetry
functions. As can be seen from the plot to the bottom right in
\fig{fig:tasy.dphitt}, the more violent radiation emerging from the
colour dipoles spanned with backward-moving top quarks leads to a
negative asymmetry over a wide range of angles, except for very large
angles where the asymmetry turns positive as a result of the Sudakov
effect. While for most predictions the cross-over occurs at
$\Delta\phi\approx3$, two results deviate considerably from the fairly
constant behaviour of the asymmetry ($A_\mathrm{FB}\sim-0.1$) for
$\Delta\phi<2$. The \Pythia~6 tune D6T and \Pythia~8 mark these two very
different ends of the low-$\Delta\phi$\/ asymmetry spectrum (with
values of $-70$\% and $+20$\%, respectively). For the former, soft
colour coherence effects are overestimated whereas, for the latter,
coherence effects have not been implemented, in
particular the initial--final $qt$\/ and $\bar q\bar t$\/ dipoles are
not yet treated as such.

\FIGURE[!t]{
  \centering
  \centerline{\hspace*{-7mm}
    \includegraphics[width=0.492\columnwidth]{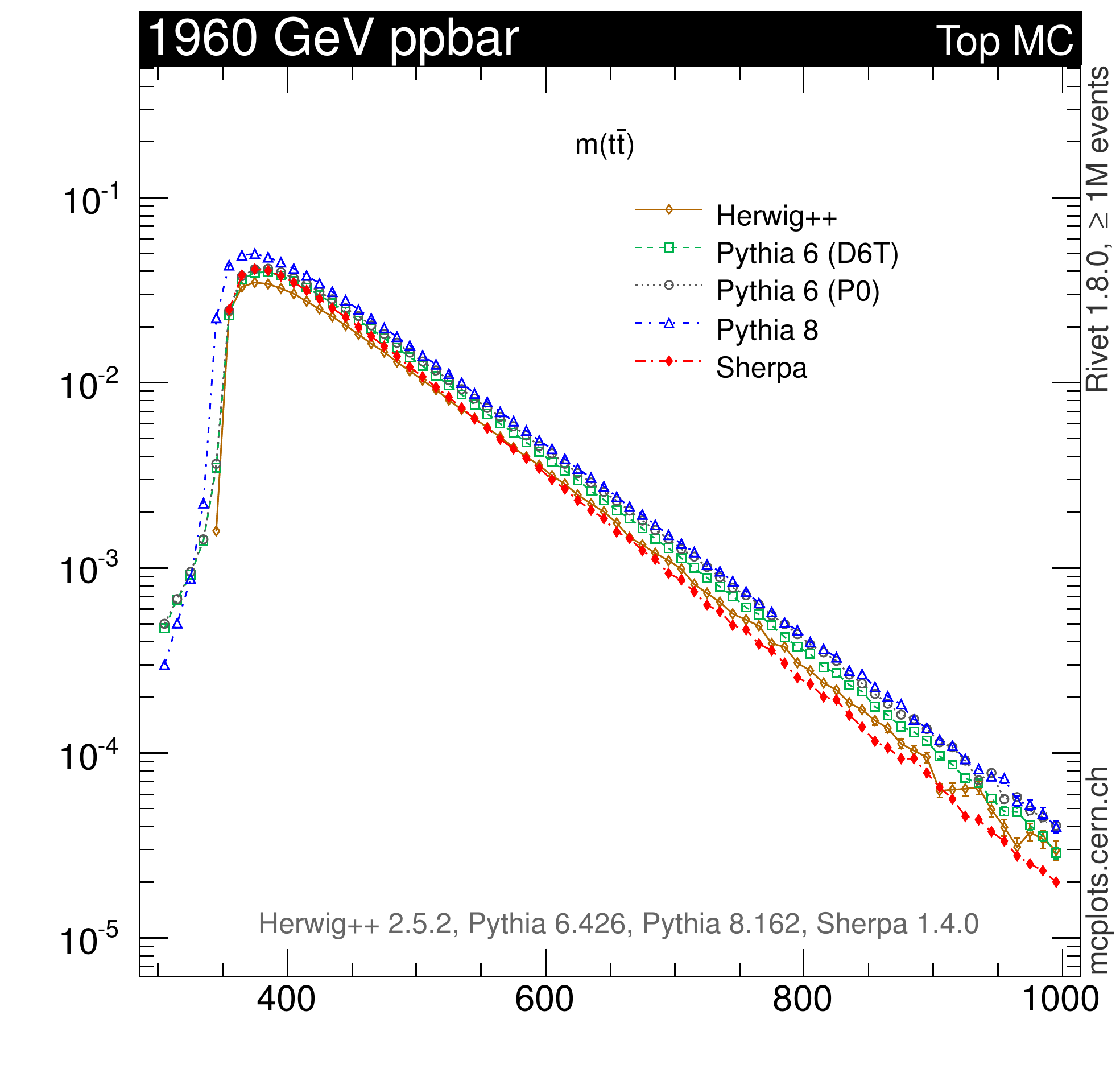}
    \includegraphics[width=0.492\columnwidth]{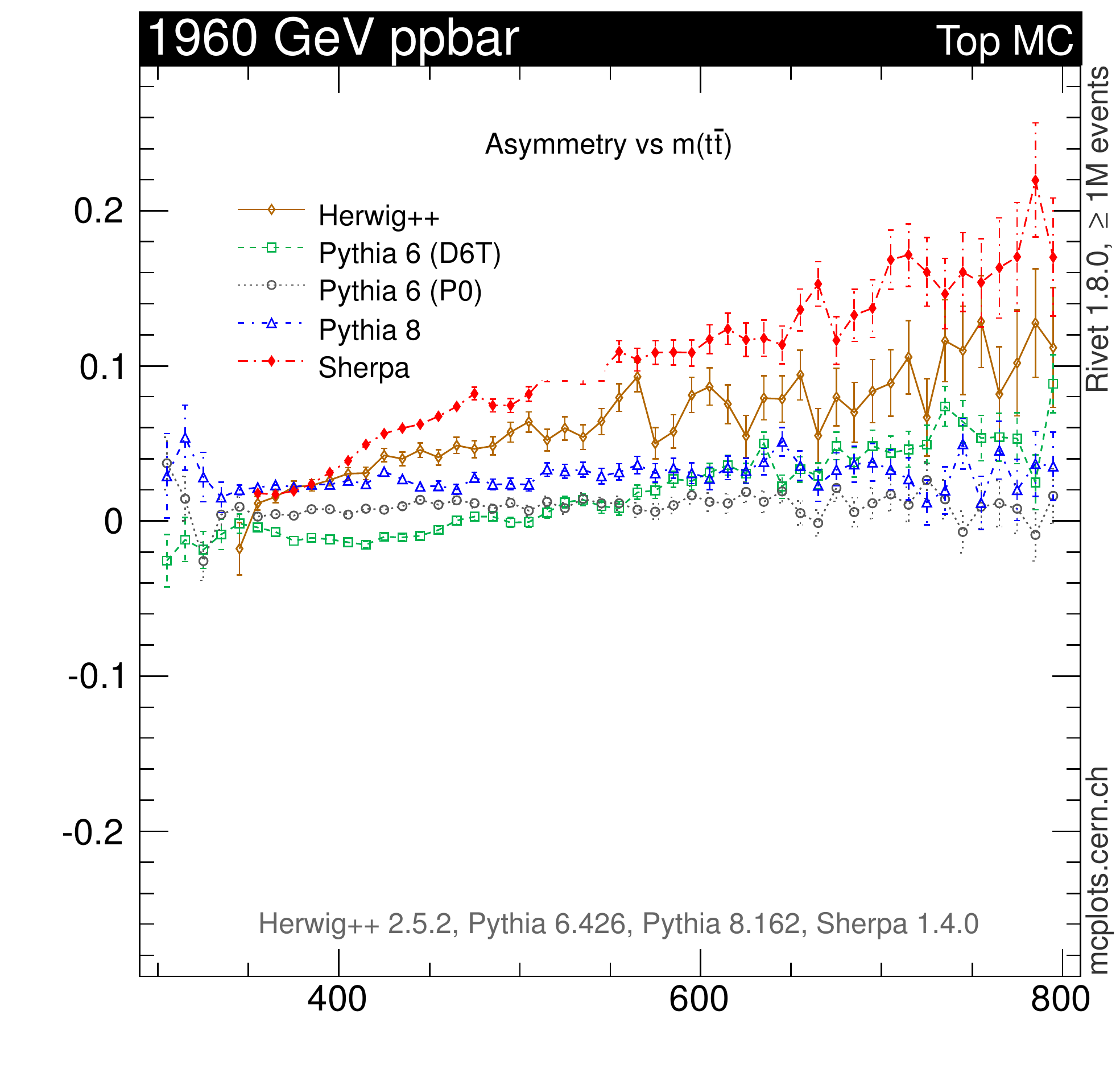}}
  \centerline{\hspace*{-7mm}
    \includegraphics[width=0.492\columnwidth]{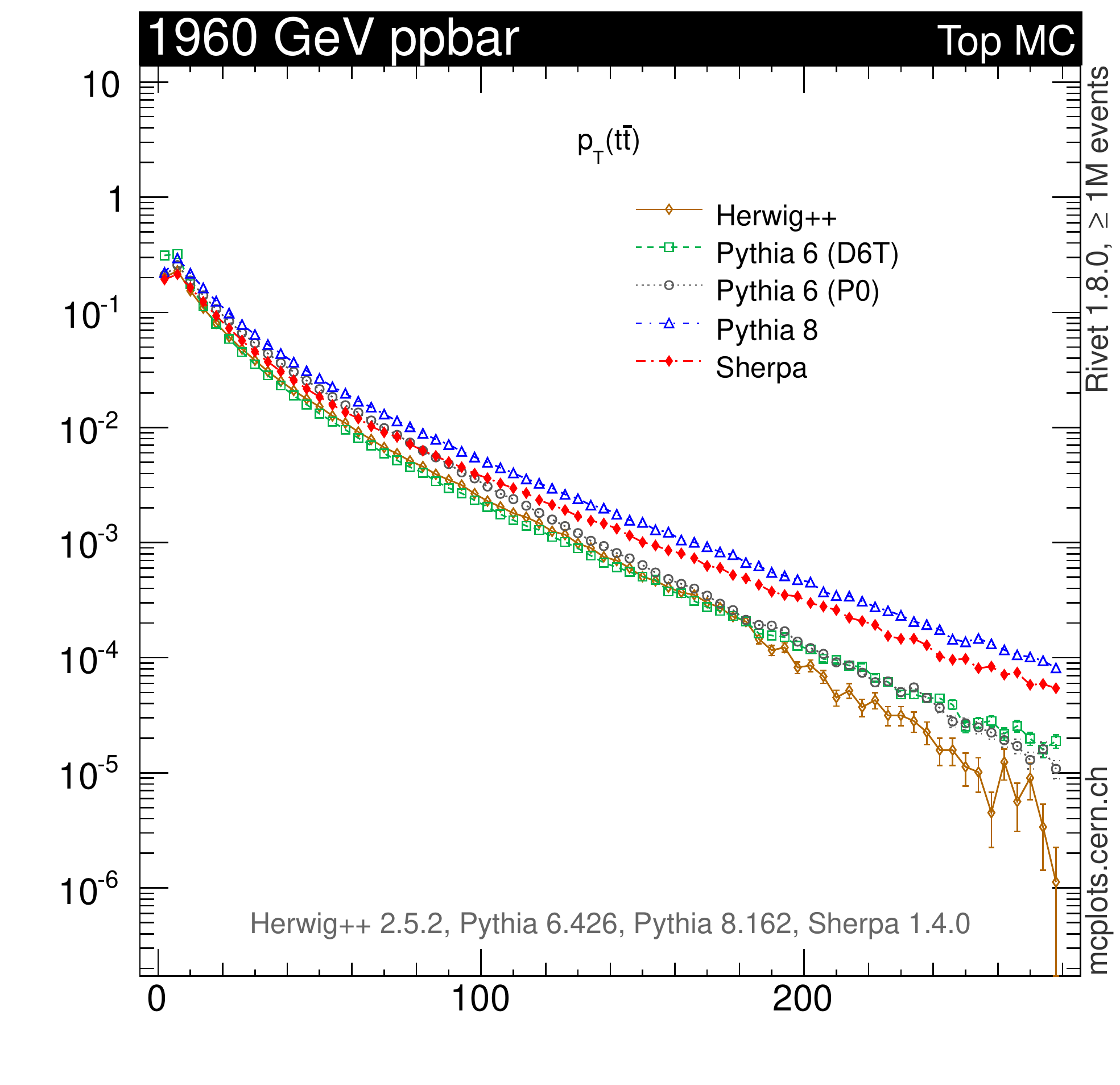}
    \includegraphics[width=0.492\columnwidth]{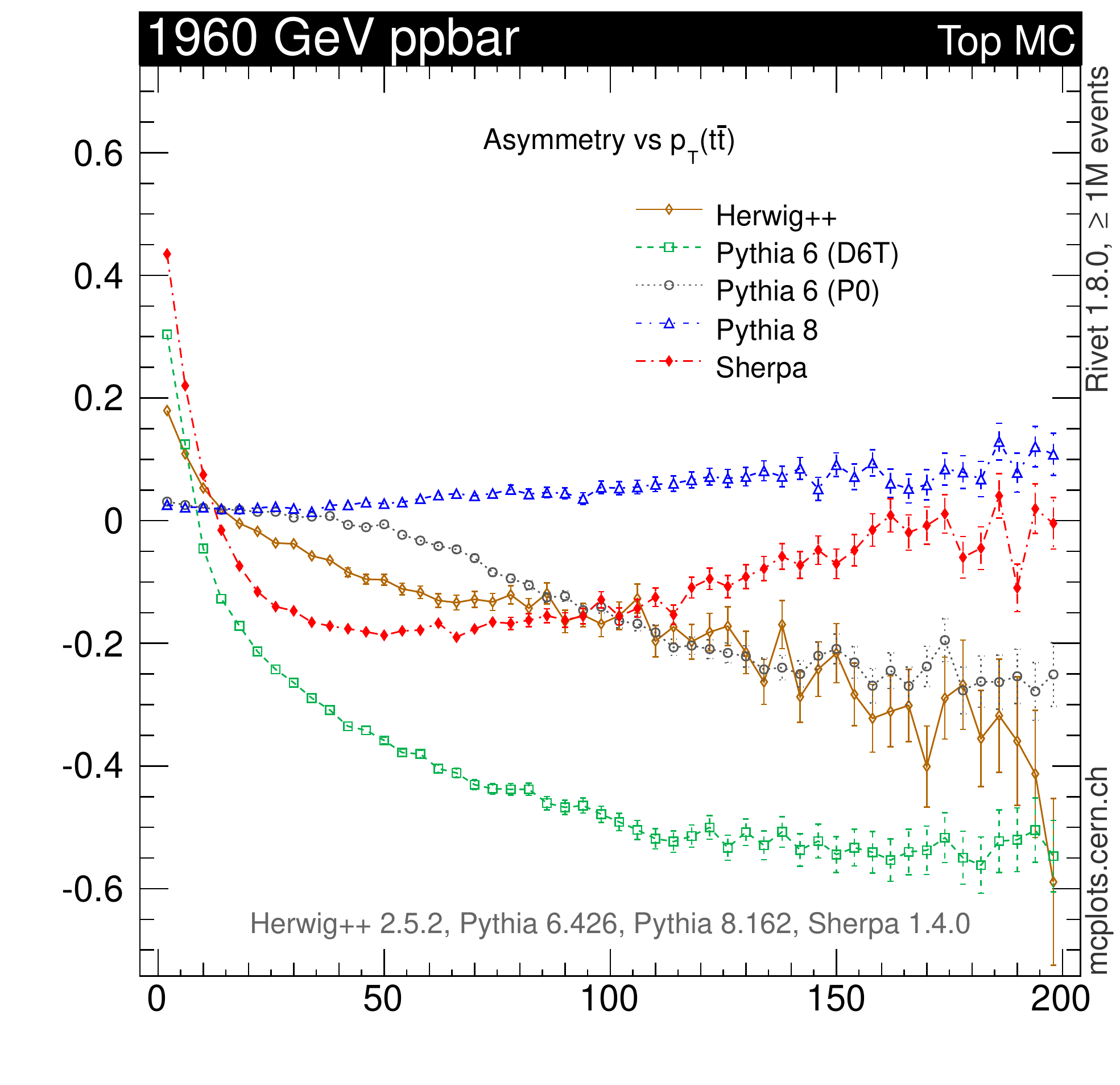}}
  \vspace*{-25pt}
  \caption{\label{fig:tasy.ptt.mtt}%
    Differential cross section (in pb/GeV) and top quark forward--backward
    asymmetry $A_\mathrm{FB}$ as a function of the mass $m_{t\bar t}$
    (upper row, in GeV) and transverse momentum $p_{T,t\bar t}$ (lower
    row, in GeV) of the top--antitop pair. Various event generator
    predictions are compared with each other, and all errors shown are
    statistical only.}
}

\fig{fig:tasy.ptt.mtt} displays the dependence of the differential
cross section and asymmetry on the mass
$m_{t\bar t}$ and transverse momentum $p_{T,t\bar t}$ of the $t\bar
t$\/ system. As for $|\Delta y|$, the Sudakov region generating
positive asymmetry contributions due to soft colour coherence
($\Delta_+>\Delta_-$) applies over the entire range of the pair
mass. We can use the following equation to better understand
this behaviour and the pair mass dependence of $A_\mathrm{FB}$:
\beq\label{eq:mtt2}
m^2_{t\bar t}\;=\;m^2_t+m^2_{\bar t}+
                  2\,E_{T,t}E_{T,\bar t}\,\cosh\Delta y-
                  2\,p_{T,t}p_{T,\bar t}\,\cos\Delta\phi
\eeq
where $E^2_T=m^2+p^2_T$ and $\Delta\phi$ is the azimuthal angle between
$\vec p_{T,t}$ and $\vec p_{T,\bar t}$. It is sufficient to focus
on the $\cosh\Delta y$\/ and $\cos\Delta\phi$\/ dependence of
$m^2_{t\bar t}$. The $\cosh\Delta y$\/ term
is forward--backward symmetric and the squared mass increases with
larger absolute rapidity differences. The $\cos\Delta\phi$\/ term in
\eq{eq:mtt2} may however reduce $m^2_{t\bar t}$, but in the hard
region only. Consequently, the $\cosh\Delta y$\/ dependence of
$A_\mathrm{FB}$ directly translates into a similar mass dependence.
Neglecting this for a moment, we also notice that for given $\Delta
y$, it is cheaper to shift the masses to larger values. This is
because of the enhancement of soft emissions ($\Delta\phi\approx\pi$)
causing an overall plus sign in the $\cos\Delta\phi$\/ term. The
imbalance in the soft-emission rate generated by soft colour coherence
between the forward and backward region thus produces harder mass
spectra in the forward region. Taken together with the $|\Delta y|$
dependence of the asymmetry, we conclude that the $\cos\Delta\phi$\/
term induces an additional growth of the asymmetry with increasing
pair mass and a small suppression in the low mass region. In cases
where the $|\Delta y|$ dependence of the asymmetry is almost zero, the
same mechanism may generate slightly negative asymmetries at low mass.

Looking at the different generator results displayed in
\fig{fig:tasy.ptt.mtt}, the $d\sigma/dm_{t\bar t}$ spectra can be seen
to differ less than the $p_{T,t\bar t}$ spectra. Again, \Pythia~8 gives
the hardest distributions. As expected, we find that the $t\bar t$\/
mass dependence of the asymmetry is determined predominantly by the
asymmetric behaviour present in $|\Delta y|$. The pattern shown on
the upper right of \fig{fig:tasy.dphitt} repeats itself here. We also
observe that small negative asymmetries are possible for low pair
masses, e.g.\ as shown for the \Pythia~6 tune D6T.

Finally, on the lower right of \fig{fig:tasy.ptt.mtt} one finds the
results obtained by the different generators for the asymmetry plotted
as a function of $p_{T,t\bar t}$. We already discussed the
characteristics of $A_\mathrm{FB}(p_{T,t\bar t})$ throughout preceding
sections; thus, recalling the discussion of the $\Delta\phi$\/ case,
the $p_{T,t\bar t}$ results are as expected and their
interpretation is straightforward. Exhibiting the asymmetry in terms
of $p_{T,t\bar t}$ naturally allows for a better discrimination in the
hard region. We thus observe that \Herwig++, \Pythia~6 P0 and \Sherpa\
differ in their description of the high-$p_T$ tail of the asymmetry,
even though they sufficiently agree in the low-$\Delta\phi$ region.
\Sherpa\ predicts a slope change around $50$ GeV indicating a possible
return in $A_\mathrm{FB}(p_{T,t\bar t})$ to zero for large $p_T$, as
seen in LO (\figs{fig:FbetapT} and \ref{fig:LO_pt_asy}). In contrast,
\Herwig++ and \Pythia~6 P0 maintain their trend towards more
negative asymmetries. At the other end of the $p_T$ spectrum,
the rise of the asymmetry towards lower $p_T$ is not shown by \Pythia~6
D6T and \Pythia~8, which is compatible with the findings for
$\Delta\phi\approx\pi$\/ in \fig{fig:tasy.dphitt}.

\bigskip
We have studied more observables than we are able to present here, so
for a more comprehensive comparison we refer the interested reader to
the corresponding {\em ``Top quark (MC)''} web-pages available under
\texttt{mcplots.cern.ch}~\cite{mcplots}. Among other things one can
find, for example, the distributions and forward--backward asymmetries
of the transverse momentum of the top quark, $p_{T,t}$, or the
rapidity $y_{t\bar t}$ of the $t\bar t$\/ pair; in addition many
observables are shown separately for the high and low pair mass or
transverse-momentum region obtained by cutting on $m_{t\bar t}$ at
$450\ \mathrm{GeV}$ or $p_{T,t\bar t}$ at $50\ \mathrm{GeV}$,
respectively.

\paragraph{Summary.}

We can say that the \Herwig++ and \Sherpa\ predictions
agree fairly well with each other, and compare quite nicely -- on a
qualitative level -- to the $A_\mathrm{FB}(m_{t\bar t})$ and
$A_\mathrm{FB}(|\Delta y|)$ results given in
Ref.~\cite{Ahrens:2011uf}. Both models incorporate soft colour
coherence on a compatible level, which consequently may be interpreted
as the source of the agreement; the differences lie in details such as
the treatment of recoils, the shower variables and the form of the
splitting functions used. These differences cause deviations in the
high-$p_T$ asymmetry spectra. The dependence on recoil effects in
\Sherpa\ is studied in more detail in \sec{sec:css.vary} below. 
In \Pythia, soft colour coherence is accounted for on a more
approximate level. Although the P0 tune of \Pythia~6 is similar to
\Herwig++ in the hard-emission domain, it differs in the description
of the Sudakov region, yielding a milder mass and $|\Delta y|$
dependence of the asymmetry. The \Pythia~6 D6T tune and \Pythia~8
exhibit larger differences. The dependence on the shower modelling in
\Pythia~6 is studied in more detail in \sec{sec:pythia.tune} below.

\subsubsection{Dependence on recoil effects: \Sherpa's \Css}
\label{sec:css.vary}
We have argued that recoil effects play an important role in producing
the asymmetries generated in coherent parton or dipole showering.
To illuminate the mechanism further, we have conducted a small case
study based on results obtained with \Sherpa's \Css. Some of the
details have been postponed to \app{app:addmat} in order to maintain
the flow of the main part.

\paragraph{Asymmetry enhancing longitudinal recoil effects.}

We have identified the unequal Sudakov form factors in forward and
backward top production as a major source of asymmetry.
The Sudakov imbalance
($\Delta_+>\Delta_-$) emerges as a result of soft colour coherence.
In addition, based on \eq{eq:incl2} we have seen that any net
migration of the type $P_{-+}>P_{+-}\ge0$, from the backward to the
forward $\Delta y$\/ phase space, leads to a positive inclusive
asymmetry and an amplification of the Sudakov or coherence effect. We
now want to trace the origin of the migration process.

In a simple dipole picture where a gluon emission stretches
(further opens the initial angle of) the starting initial--final
$qt$\/ or $\bar q\bar t$\/ dipole, one can easily account for
$\Delta y=\Delta\tilde y+\epsilon$\/ on average.\footnote{In
  \fig{fig:ttbar_coh}~(a)~and~(b), this corresponds to a reduction of
  the scattering angle, i.e.\ a kick of the (anti)top quark in the
  incoming (anti)quark direction, thereby mitigating the acceleration
  of colour charges.} Here we denote the top quark rapidity difference
before the emission (or at the LO generation level) by
$\Delta\tilde y$, and $\epsilon>0$ expresses a small positive shift.
Using a simple generation cut on $\Delta\tilde y$, we can test this
hypothesis by counting and analyzing the events that end up in the
backward/forward region after exclusively showering $t\bar t$\/ events
produced at LO under the constraint $\pm\Delta\tilde y>0$.

\FIGURE[!t]{
  \centering
  \centerline{
    \includegraphics[width=0.6\columnwidth]{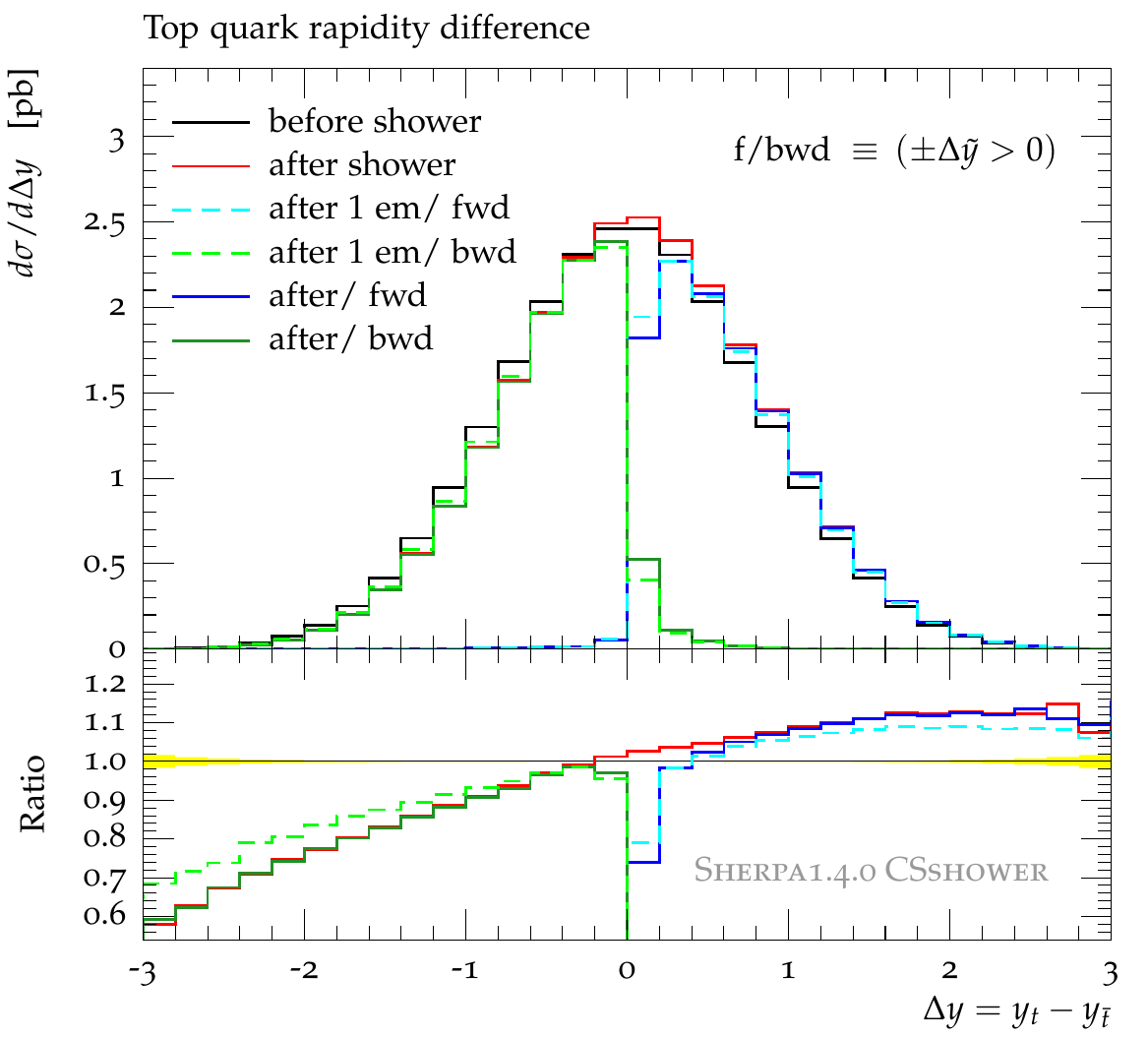}}
  \centerline{
    \includegraphics[width=0.6\columnwidth]{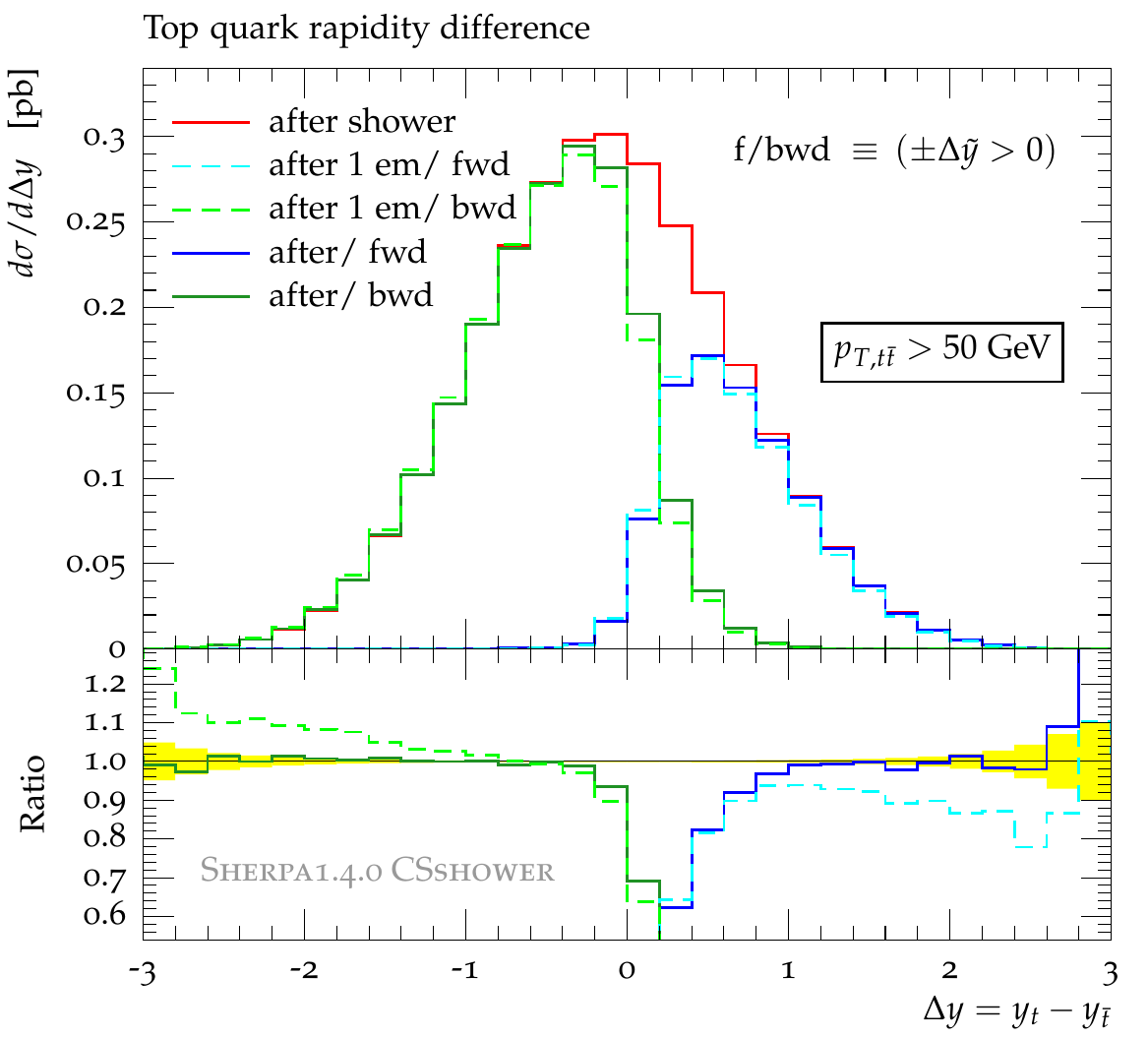}}
  \vspace*{-17pt}
  \caption{\label{fig:migrate}%
    The $\Delta y$\/ distributions for different LO generation and
    shower modes as predicted by \Sherpa's \Css. Dashed lines
    correspond to results taken from one-emission (``1~em'') showers,
    while solid lines depict those after complete showering, except
    for the black curve, which depicts the fixed LO parton-level
    prediction in the upper subfigure. Blue [green] lines show the
    outcomes when the LO $t\bar t$\/ phase-space generation is
    constrained to the forward (``fwd'', $\Delta\tilde y>0$)
    [backward] region. Each lower part contains a ratio plot using the
    respective list's first prediction as reference.}
}

\FIGURE[!t]{
  \centering
  \centerline{
    \includegraphics[width=0.6\columnwidth]{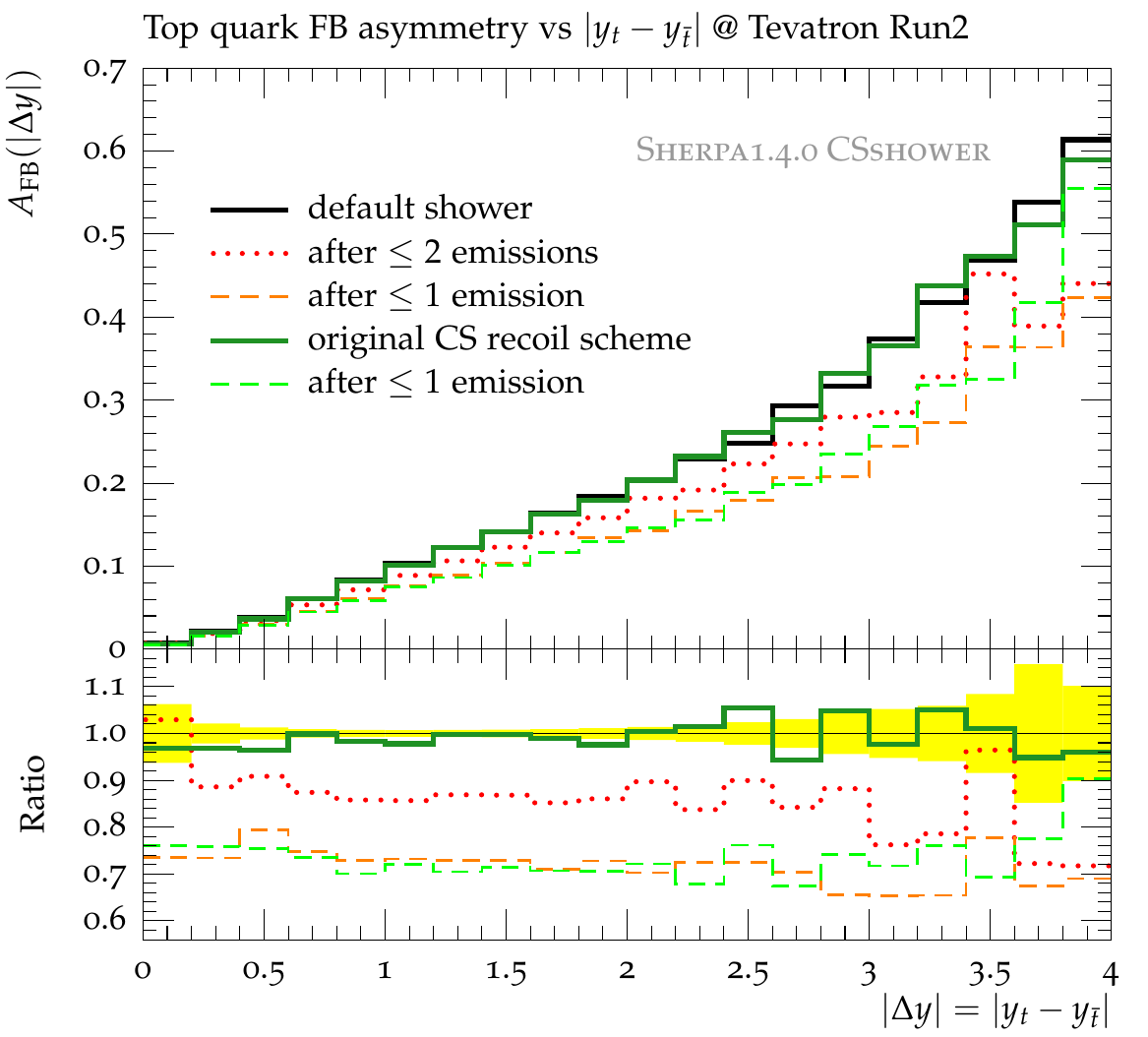}}
  \centerline{
    \includegraphics[width=0.6\columnwidth]{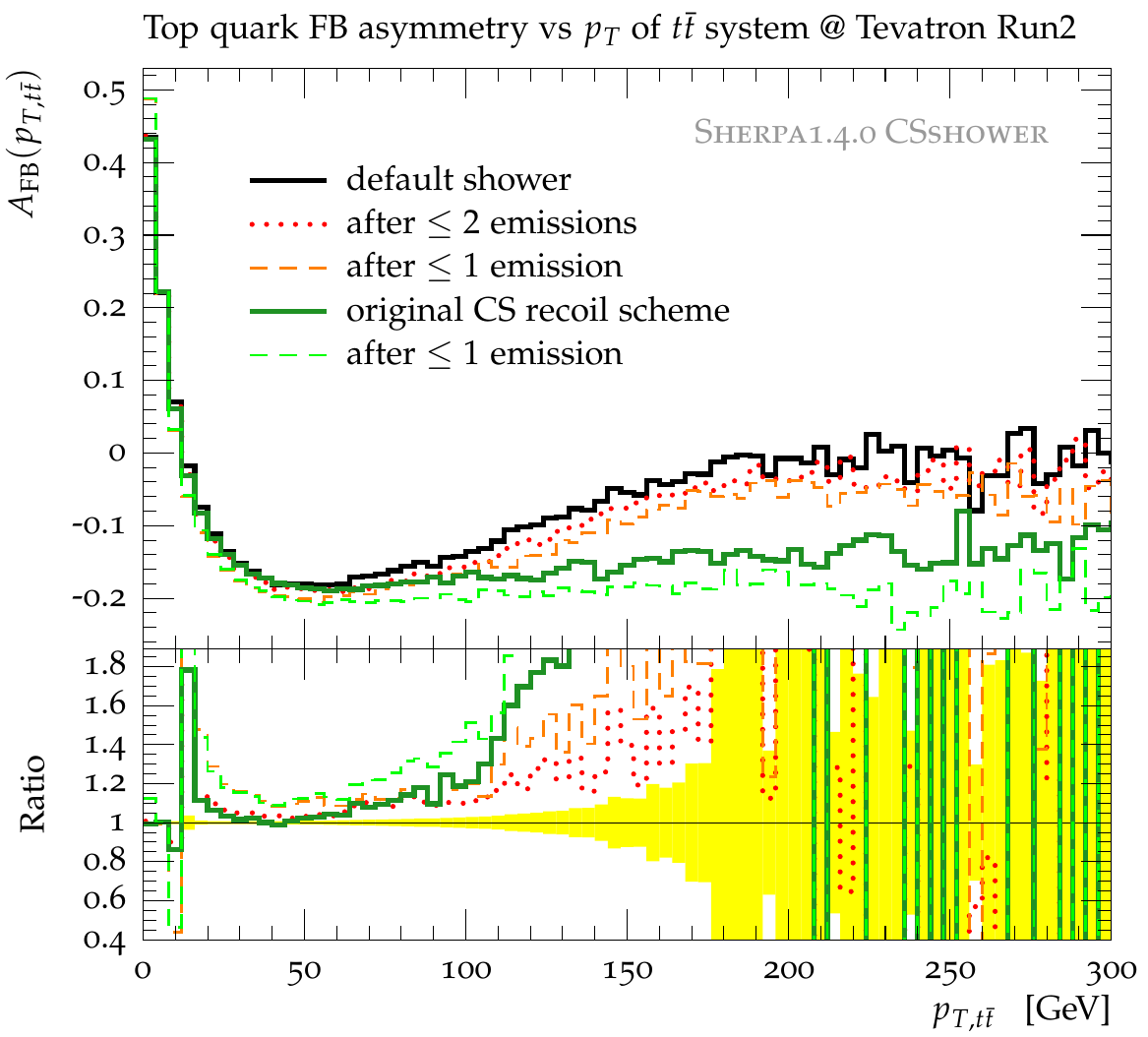}}
  \vspace*{-17pt}
  \caption{\label{fig:recoils}%
    The forward--backward asymmetries versus $|\Delta y|$ (upper
    panel) and $p_{T,t\bar t}$ (lower panel) for the two different
    recoil strategies available for the \Css\ implementation in
    \Sherpa. Results gained by the corresponding one(two)-emission(s)
    showers are depicted as well. Each subfigure is supplemented by a
    ratio plot using the default \Css\ prediction for reference.}
}

\fig{fig:migrate} shows the main results of this migration test; more
details are compiled in \app{app:addmat}. We have plotted the outcomes
of $t\bar t$\/ production at leading, fixed order (labelled ``before
shower'') and of several \Css\ runs (all labelled ``after\ ...'')
grouped according to unconstrained, $\Delta\tilde y>0$ (``fwd'')
and $\Delta\tilde y<0$ (``bwd'') LO phase-space generation: we
distinguish between completely showered runs and runs where showering
was terminated after just one emission. For the latter, we only show
the results obtained with the restricted LO phase space. Focusing on
$\Delta y$\/ distributions, we make a number of observations, broadly
confirming the physics of the simple dipole picture suggested above:
\begin{itemize}
\item showering generates a positive asymmetry, which is growing with
  larger $|\Delta y|$ (compare ``before'' and ``after shower''
  results).\\[-7mm]
\item migrations are small, happen locally but yet across the entire
  $\Delta y$\/ phase space; the $-\to+$ direction and, therefore,
  $-\to+$ cross-overs are favoured: (1)~the ``bwd'' generated results
  extend into the $\Delta y>0$ domain filling the deficit close to the
  transition region left by ``fwd'' generated events that now populate
  larger $\Delta y$, and (2)~the migration processes in the opposite
  direction are largely suppressed.\\[-7mm]
\item the largest effect already originates from the first emission,
  the gluon emission of the initial--final $qt$\/ and $\bar q\bar t$\/
  dipoles (compare dashed with solid lines).\footnote{This is also
    reflected by finding comparable $A_\mathrm{FB}$ when running the
    \Css\ in one-emission mode ($4.1$\%, $2.7$\%, $6.6$\%, $6.9$\% and
    $-17.1$\%, to be compared to the \Sherpa\ 1.4.0 entries of
    \tab{tab:totasym}, in this order).} This does not necessarily mean
  that the corrections from multiple emissions are negligible; they
  easily give 10--20\% effects, as can be seen in both
  \figs{fig:migrate} and \ref{fig:recoils}.
\end{itemize}
This pattern carries over to the high-$p_T$ region,
$p_{T,t\bar t}>50$~GeV, except for the fact that the total asymmetry
now turns negative, cf.\ the lower panel of \fig{fig:migrate}. The
migration is more severe, but cannot overcome the overall negative
trend, caused by the more violently radiating ``bwd'' generated
initial dipole configurations: the radiation imbalance predominates
over the migration effect.

\paragraph{Comparison of recoil strategies.}

\Sherpa's \Css\ implementation provides two recoil schemes, the
default one, see Refs.~\cite{Platzer:2009jq,Hoeche:2009xc} and the
original CS scheme as advocated in
Refs.~\cite{Catani:1996vz,Schumann:2007mg,Winter:2007ye}.
They differ mainly in their treatment of the transverse recoils. The
distribution of the longitudinal recoil momenta effectively is the
same in both schemes, as documented in the top panel of
\fig{fig:recoils} where we show $A_\mathrm{FB}$ as a function of
$|\Delta y|$ (exhibiting all of the characteristics described
earlier).\footnote{The $A_\mathrm{FB}(m_{t\bar t})$ are also broadly
  unaffected by the scheme change; similarly the
  $A^{\mathrm{(cut)}}_\mathrm{FB}$ deviate only marginally from the
  default \Sherpa\ 1.4.0 numbers stated in \tab{tab:totasym}: $5.4$\%,
  $3.1$\%, $9.7$\%, $8.3$\% and $-17.8$\%, to be compared in this
  order.}
We see again, the bulk of the asymmetry is already produced by the
one(two)-emission(s) showers, similarly for $A_\mathrm{FB}(p_{T,t\bar t})$
displayed in the lower panel of \fig{fig:recoils}. The $p_T$-dependent
asymmetry function is the prototype of distributions discriminating
clearly between the two recoil strategies: the original CS scheme is
more like that in the Catani--Seymour NLO calculational
scheme~\cite{Catani:1996vz}: the~(transverse) recoil from a
gluon emitted off a $qt$\/ configuration (and likewise
off a $\bar q\bar t$\/ one) is compensated by the top quark,
regardless of its role in the emission process (emitter or spectator).
The prediction given by the original scheme therefore
remains flat at about $-15$\% for high $p_{T,t\bar t}$ while the
default prediction levels off close to zero from below. In the
default scheme, the gluon recoil is rather divided over the
entire set of final-state partons.
This requires an additional transverse boost combined with a rotation
that in turn washes out the radiation imbalance between the forward
and backward regions for very large $p_{T,t\bar t}$.

\subsubsection{Dependence on shower model: \Pythia}\label{sec:pythia.tune}

For processes like $t\bar{t}$\/ production, which do not contain any QCD
jets at the Born level, both \Pythia~6 and \Pythia~8 use 
so-called ``power showers'' \cite{Plehn:2005cq} to populate the $t\bar{t}j$\/
phase space. Since the LL splitting kernels generally represent an
overestimate in the region of very hard jets, a factor that suppresses
such emissions has been introduced in the $p_\perp$-ordered showers
in both \Pythia~6  and \Pythia~8, similar in spirit to a matrix-element
correction \cite{Miu:1998ju} but with a much simpler analytical structure.
In \Pythia~8, the suppression factor is derived from universal
$t$-channel arguments \cite{Corke:2010zj} and does not depend on the colour
structure of the event, wherefore it does not contribute to the
generation of any $t\bar{t}$\/ asymmetry. In \Pythia~6, the
suppression factor is \cite{Skands:2010ak}
\begin{equation}
P_{\mathrm{accept}}\;=\;\min\left\{1,~P_{67} ~ \frac{s_{D}}{4\,p_{T{\mathrm{evol}}}^2}\right\}~,
\label{eq:p67}
\end{equation}
where $P_{67}$ corresponds to the parameter \texttt{PARP(67)} in the
code, $p_{T\mathrm{evol}}$ is the evolution scale for the branching, 
and $s_D$ is the invariant mass squared of the radiating parton
with its colour partner, with all momenta crossed into the final
state (i.e.\ it is $\hat{s}$\/ for annihilation-type colour flows and
$-\hat{t}$\/ for an initial--final connection). This is motivated
partly from studies of similar factors in the context of ``smooth
ordering'', introduced in \cite{Giele:2011cb}. Finally, in the
$Q^2$-ordered shower model in \Pythia~6~\cite{Sjostrand:2006za}, 
a veto on the emission angle is placed, which depends explicitly on
the direction of the colour partner.

To illustrate the effect of these choices on the asymmetry, we show
the dependence of the asymmetry on $p_{T,t\bar{t}}$ for 
five \Pythia~6 tunes in \fig{fig:tasy.tunes.ptt}. 

\FIGURE[!t]{
  \centering\centerline{\hspace*{-7mm}
    \includegraphics[width=0.492\columnwidth]{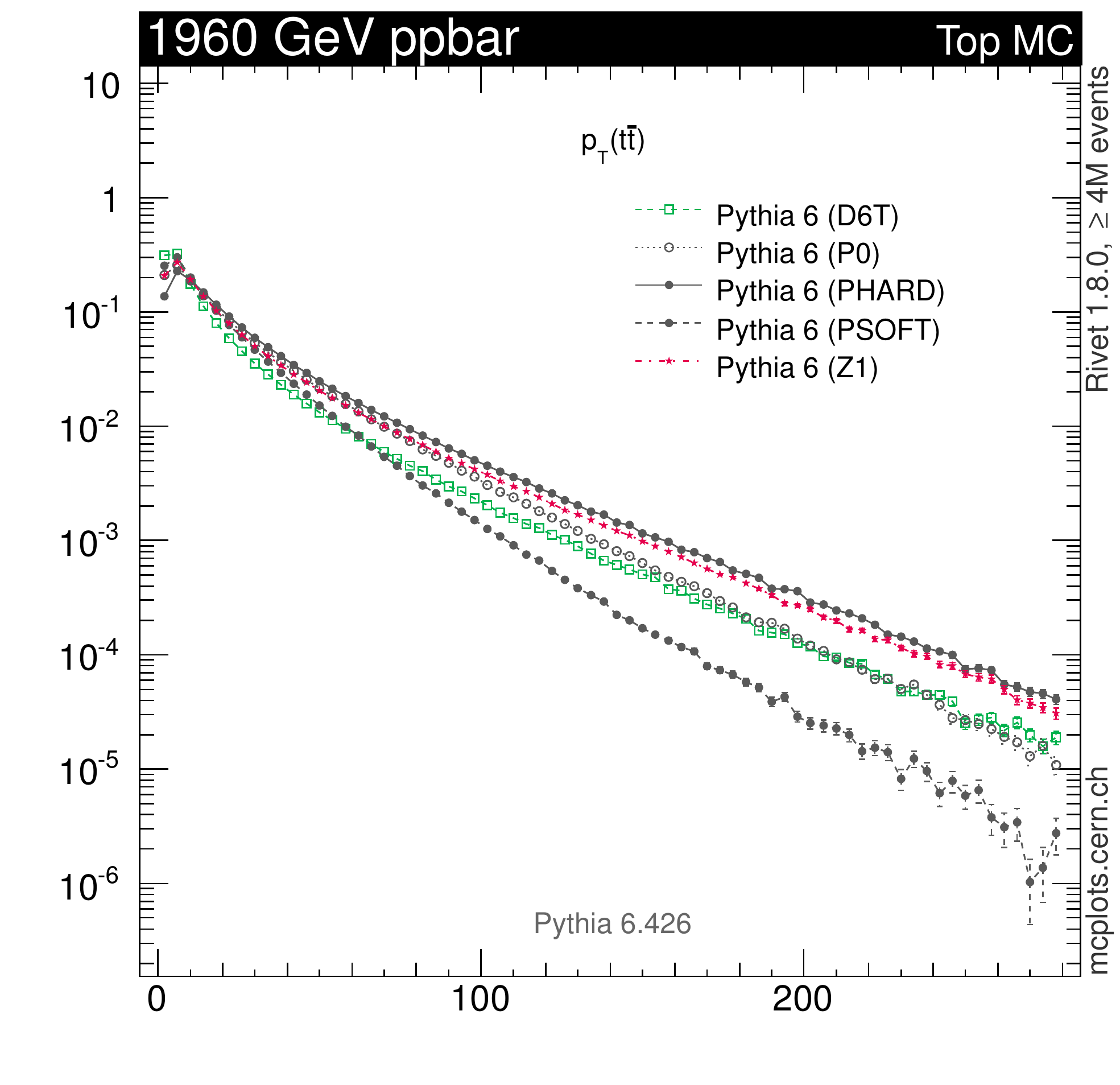}
    \includegraphics[width=0.492\columnwidth]{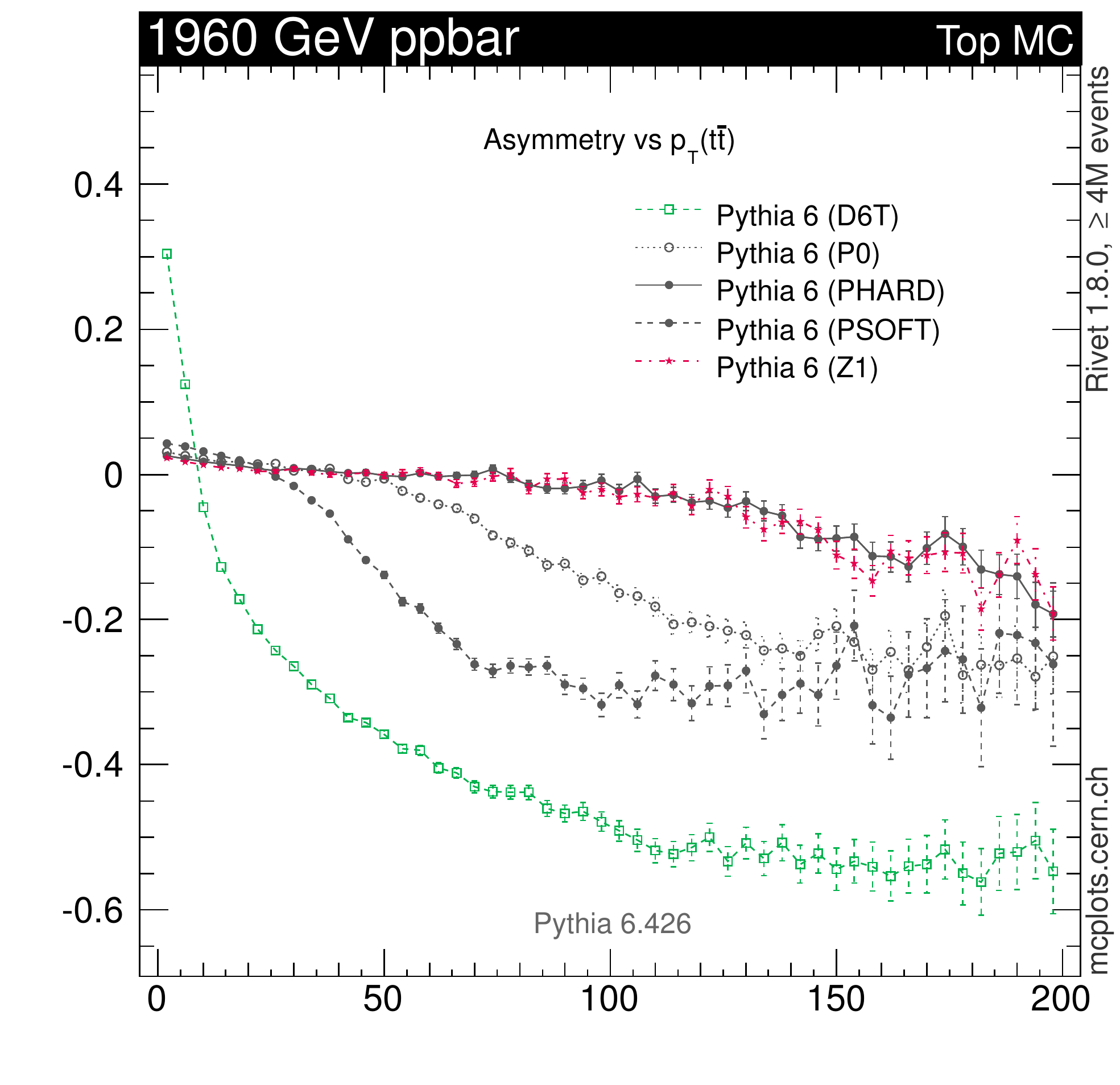}}
  \vspace*{-25pt}
  \caption{\label{fig:tasy.tunes.ptt}%
    Transverse-momentum distribution $d\sigma/dp_{T,t\bar t}$ (in
    pb/GeV) of the top quark pair (left) and the related top quark
    forward--backward asymmetry (right) in five different tunes of
    \Pythia~6.}
}

The $Q^2$-ordered D6T tune includes the explicit angular cut on the
emission angle mentioned above, which a priori should produce an effect
qualitatively similar to that of the angular-ordered showers in
\Herwig++. However, in \Pythia's $Q^2$-ordered shower, the 
effect is amplified, as follows. When an
initially massless ISR parton evolves to become a massive
jet, its virtuality is generated by reducing its momentum while
keeping its energy unchanged. The momentum of whatever the ISR parton is
recoiling against (here the $t\bar{t}$\/ system) is then also reduced
to ensure momentum conservation. Thus, to get a $t\bar{t}$\/ pair with a
certain $p_T$, the $Q^2$-ordered shower must first radiate a massless
ISR parton with an initially  
much larger value of $p_T$, which is then reduced when that parton
acquires a virtuality.  The fact that this compresses
the $p_T$ spectrum is well-known from the Drell--Yan case (cf.\
e.g.~\cite{mcplots}) and can be counteracted by choosing a low
renormalization scale for $\as$ (as has been done in D6T, which
uses $\mu_\mathrm{R}=0.45\,p_T$), thus
bringing the $p_T$ spectrum itself back into rough agreement with
other models, as illustrated in the left-hand panel of
\fig{fig:tasy.tunes.ptt}. The asymmetry spectrum, however, remains
compressed, as shown in the right-hand panel. 

Among the four $p_\perp$-ordered tunes, the Perugia 0 (P0), Perugia HARD
(PHARD), and Perugia SOFT (PSOFT) ones \cite{Skands:2010ak} use
the suppression factor defined by \eq{eq:p67}, while Z1 does not apply 
any suppression. The central Perugia 0 (P0) tune uses $P_{67} = 1$. 
This generates an asymmetry that
begins to turn on at roughly $p_T = m_t/2$.  The HARD and SOFT variations 
use $P_{67}=4$ and $P_{67}=1/4$, respectively, which modifies the
turn-on point. Neither the HARD nor the Z1 tunes exhibit any
significant asymmetries in the region plotted here, similarly to the
case in \Pythia~8.

\section{Summary and implications}\label{sec:conclude}
The studies presented above arose from the initially surprising
observation that Monte Carlo event generators can produce non-zero
forward--backward asymmetries in top pair production, even when treating
the relevant subprocess $q\bar q\to t\bar t$\/ at leading order, which
has no such asymmetry.  Our studies show that these asymmetries
arise from valid physics built into generators with coherent parton or
dipole showering.  While not quantitatively correct in every detail,
the coherent showering approximation captures essential features
of the physics not hitherto well understood, which may serve as a guide to
the contributions of higher orders.

The generated asymmetries are of two kinds.  First, in the
differential cross section at non-zero transverse momentum of the
top pair, a negative asymmetry results from the extra QCD radiation
emitted when the top quark is produced backwards in the rest frame of
the pair.  This effect is manifest in the $q\bar q\to t\bar t g$\/
matrix element and is present in the generators in the soft
approximation, with a colour coefficient that is exact in the
large-$N$\/ limit but 60\% too large at $N=3$.

In fixed-order perturbation theory, the asymmetry at non-zero $p_T$ of
the pair tends to zero from below as $p_T\to 0$.  However, precisely at
$p_T=0$ there are singular virtual contributions that lead to a
positive overall inclusive asymmetry which grows with increasing
invariant mass of the pair.  The event generators perform an
approximate all-order resummation of perturbation theory, which
smears out the singular contributions at $p_T=0$ over a finite
Sudakov region, and so the asymmetry changes sign at some point
and becomes positive at small $p_T$. The precise switching point is
sensitive to finite terms and higher-order corrections not included in
the generators, but the change of sign is a striking general prediction
that should be investigated experimentally.\footnote{For a related
  quantity, the asymmetry when there is an extra jet with
  $p_T>20$~GeV, a negative value has indeed been reported~\cite{Abazov:2011rq}.}
It is also worth noting that, because of this switch, a bias towards
low $p_T$, or against extra jet production, in the method used 
to reconstruct the tops could lead to a significant upward shift in
the measured inclusive asymmetry.

The other type of generated asymmetry is an overall inclusive one,
positive and growing in value with increasing invariant mass of the
pair.  In fixed-order perturbation theory, such an asymmetry
appears at order $\as$ relative to the Born process
and is due to a positive asymmetry in the virtual correction,
which dominates over the negative contribution of real emission
discussed above.
The event generators implicitly contain virtual corrections, in the
form of the Sudakov factors that drive the showers and produce $p_T$ smearing.
These factors are a reflection of unitarity, which implies that showering
cannot change the inclusive cross section from the value established
by the primary subprocess.  One might therefore think that the
inclusive asymmetry also could not change from zero when the
primary process is symmetric.  However, that is not the case, because
the asymmetry is not an inclusive cross section and so is not protected by
unitarity.

In fact, the same effect that generates a negative
asymmetry at non-zero $p_T$, namely the extra radiation in
backward top production, tends to produce a positive inclusive
asymmetry.  As expressed by \eq{eq:incl2}, it arises from
the difference between the Sudakov factors for forward and
backward top production and the migration of recoiling
top quarks between hemispheres.  Thus it is a combination
of real and virtual effects, which is of relative order $\as$,
because the difference of Sudakov factors is of that order
while the forward and backward migration probabilities
are pure numbers, modulo higher-order corrections,
with magnitudes that depend on how the event generator treats
recoils.  The fact that recoil strategies based on colour flow
produce inclusive asymmetries, similar to the full fixed-order one,
suggests that the asymmetry can be regarded as arising in this
way. Such a viewpoint could serve as a guide towards the more correct
treatment of recoils, and conversely as an indication of the possible
effects of higher orders beyond the range of explicit
calculations.\footnote{Progress has been reported in the literature,
  see for example~\cite{Dittmaier:2008uj,Alioli:2011as,Baernreuther:2012ws}.}

We believe that these findings have important implications for the
interpretation of the experimental data.  At the very least, one needs
to be aware that the available event generators can produce significant
asymmetries where none were previously expected.  Monte Carlo
estimates of corrections to asymmetries could be affected by this,
particularly corrections to ``parton level''. Moreover, these
corrections will likely be model-dependent as documented by the
detailed parton-shower comparison presented here. The results depend
on the way colour coherence is implemented in the various codes, which
we have summarized at the end of \sec{sec:excl_comp}.

On a more theoretical
level, the fact that these asymmetries are due to recoils points to
the importance of recoil effects, which are often neglected in
estimates of higher orders based on soft gluon resummation.

There are clearly many directions in which the studies presented here
could be extended.  The effects of asymmetries produced or enhanced
by parton showering in generators that match to an NLO calculation,
such as \MCNLO\ and \POWHEG, need to be assessed.  Similarly
for schemes that match to LO multi-parton matrix elements.
Analogous effects will also be present in the $t\bar t$\/
charge asymmetry, currently being investigated at the LHC,
and in other processes where the colour flow of the primary
process affects parton showering.



\section*{Acknowledgments}

We are grateful to Anton~Karneyeu for having made the
\texttt{mcplots.cern.ch} production system available to us, which greatly
facilitated making plots for the paper. We also thank Keith~Hamilton,
Stefan~H\"oche, Frank~Krauss, Michelangelo~Mangano,
Kirill~Melnikov, Stephen~Mrenna, Stefan~Prestel, and
Gavin~Salam for useful comments and discussions.
PS and BW thank the Galileo Galilei Institute for Theoretical Physics for
hospitality and the INFN for partial support during part of this work.
BW acknowledges the support of a Leverhulme Trust Emeritus Fellowship,
and also thanks the CERN Theory Group and the Center for Cosmology and
Particle Physics at New York University for hospitality. 


\clearpage
\appendix
\section{Appendix: Additional \Sherpa\ \Css\ studies}\label{app:addmat}

\paragraph{Migration tests.}

\FIGURE[!t]{
  \centering
  \includegraphics[width=0.484\columnwidth]{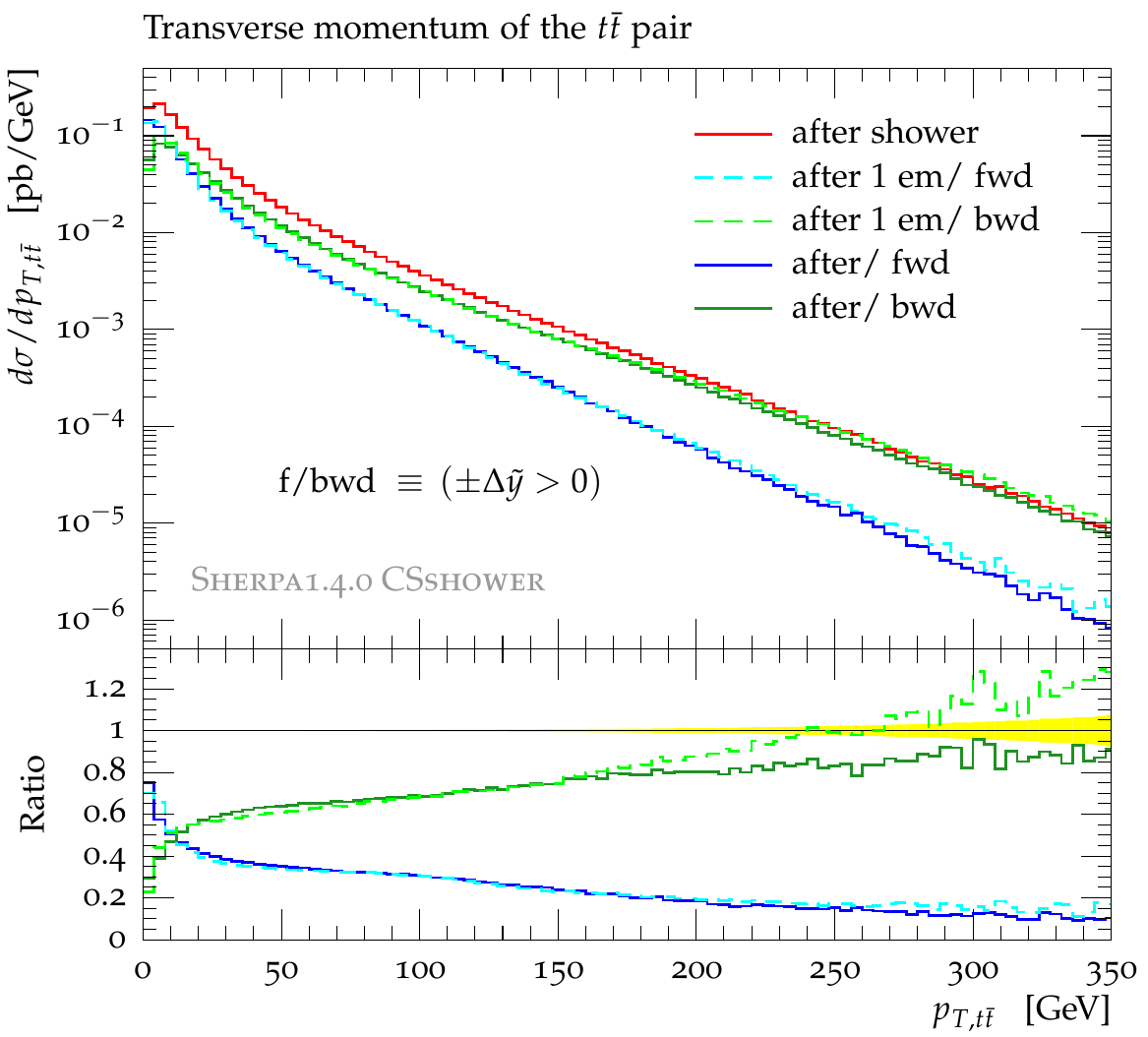}
  \includegraphics[width=0.484\columnwidth]{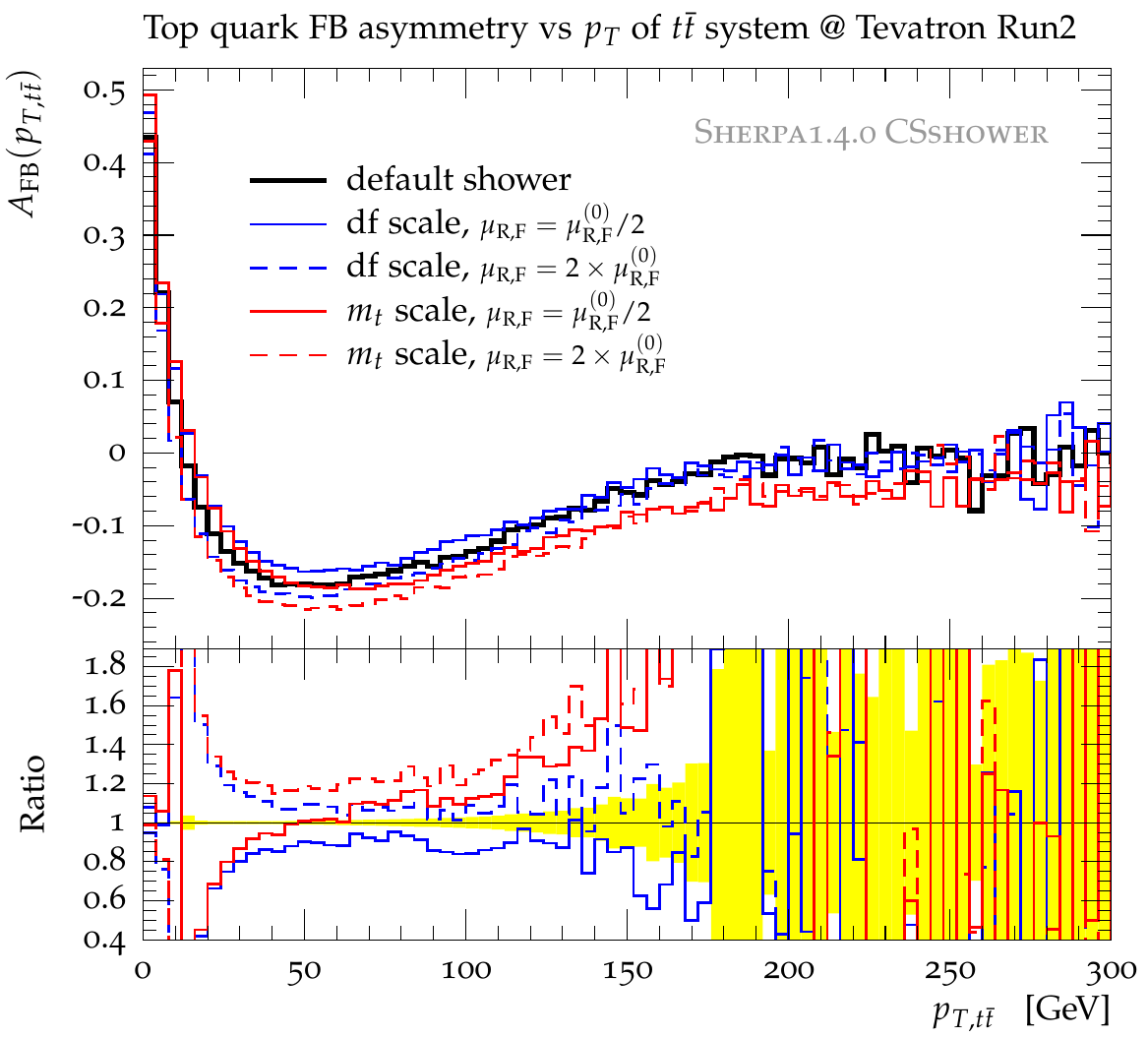}
  \caption{\label{fig:pTtt:scales}%
    (Left) \Sherpa\ \Css\ predictions for the $p_{T,t\bar t}$
    distribution using different LO generation and shower modes.
    Dashed lines correspond to results taken from one-emission
    (``1~em'') showers, while solid lines depict those after complete
    showering. Blue and green lines show the outcomes under
    constrained LO $t\bar t$\/ phase-space generation, to the forward
    (``fwd'', $\Delta\tilde y>0$) and backward region, respectively.
    (Right) \Sherpa\ \Css\ results for $A_\mathrm{FB}(p_{T,t\bar t})$
    obtained under scale choice variations ($m_{\perp,t}$-like, i.e.\
    default versus $m_t$) combined with simultaneous variation of
    $\mu_\mathrm{R}$ and $\mu_\mathrm{F}$, which was also applied to
    the parton showering.}
}

In the left part of \fig{fig:pTtt:scales} the radiation imbalance
between ``fwd'' and ``bwd'' initial dipole configurations, seen in
\fig{fig:migrate}, is clearly documented by the respective predictions
for the $p_T$ distribution of the $t\bar t$\/ pair. The LO
configurations emerging from the ``fwd'' ($\Delta\tilde y>0$) phase
space generate a more steeply falling $p_T$ spectrum with respect to
that produced by the ``bwd'' ($\Delta\tilde y<0$) generated dipoles.
The turn-over already appears at $p_{T,t\bar t}\sim10$~GeV, slightly
below the turn-over found in $A_\mathrm{FB}(p_{T,t\bar t})$, cf.\
\figs{fig:recoils} (bottom) and \ref{fig:pTtt:scales} (right). The
upward shift is a result of the migration in $\Delta y$.

\TABLE[!t]{
\centering\small
\begin{tabular}{cclcccl}\hline\hline
  \multicolumn{1}{c}{
    \rule[-3.5mm]{0mm}{9mm}~~~~~Generated~$\Delta\tilde y$~~~~~} &
  \multicolumn{1}{c}{Inclusive} &&
  \multicolumn{2}{c}{$p_{T,t\bar t}/\mathrm{GeV}<50$} &
  \multicolumn{1}{c}{$p_{T,t\bar t}/\mathrm{GeV}>50$}&\\[-1mm]
  \rule[-3.0mm]{0mm}{8mm}phase space & $r_\mathrm{fwd}$ [\%] &&
  $\varepsilon^{\mathrm{(cut)}}$ [\%] & $r^{\mathrm{(cut)}}_\mathrm{fwd}$ [\%] &
  $r^{\mathrm{(cut)}}_\mathrm{fwd}$ [\%] &\\\hline
  \rule[0mm]{0mm}{5.0mm}%
  $<0$              &$5.9$ &&$83.6$&$3.7$ &$16.7$&\\[0mm]
  $[-0.4,0]$        &$14.7$&&$87.8$&$10.2$&$46.8$&\\[0mm]
  $[-\infty,\infty]$&$52.7$&&$88.0$&$54.1$&$42.7$&\\[0mm]
  $[0,0.4]$         &$98.2$&&$90.6$&$98.5$&$95.2$&\\[0mm]
  $>0$              &$99.2$&&$92.4$&$99.3$&$97.7$&\\[1mm]\hline\hline
\end{tabular}
\caption{\label{tab:migrate}%
  Forward ($\Delta y>0$) cross section percentages,
  $r^{\mathrm{(cut)}}_\mathrm{fwd}$, according to \Sherpa's \Css,
  where $r^{\mathrm{(cut)}}_\mathrm{fwd}=
  \sigma^{\mathrm{(cut)}}\big\rfloor_{\Delta y>0}\,\big/
  \sigma^{\mathrm{(cut)}}$. The LO cross section $\sigma$\/ is
  $4.94$~pb, dropping to $0.955$~pb under the influence of the
  rapidity-difference generation cuts $0\le\pm\Delta\tilde y\le0.4$
  applied to LO $t\bar t$\/ hadroproduction. For the low-$p_T$ region,
  $p_{T,t\bar t}<50$~GeV, the cut efficiencies are stated explicitly.}
}

Another way to express the results of \fig{fig:migrate} uses the
after-showering, forward ($\Delta y>0$) cross section fractions, which
we define as $r^{\mathrm{(cut)}}_\mathrm{fwd}=
\sigma^{\mathrm{(cut)}}\big\rfloor_{\Delta y>0}\,\big/\sigma^{\mathrm{(cut)}}$.
\tab{tab:migrate} shows them for different before-showering, $t\bar t$\/
rapidity-difference regions, listed in ascending order of
$r_\mathrm{fwd}$. This is to summarize, on a more quantitative level,
all earlier findings: considerably larger migration in $-\to+$ than
opposite direction (rows 1, 2 versus 4, 5); a factor $\sim3$ increased
migration in both directions for harder emissions ($p_{T,t\bar t}>50$~GeV);
milder migration in the low-$p_T$ with respect to the high-$p_T$
region. The increasing $p_T$-veto efficiencies reflect once more the
radiation imbalance between forward and backward phase spaces.\footnote{%
  Again, the entire pattern already occurs after one emission: for the
  $\Delta\tilde y<0$ mode, the $r^{\mathrm{(cut)}}_\mathrm{fwd}$ drop
  by 1\% only. On similar note, the numbers barely change when
  switching to the other \Css\ recoil scheme.}
Focusing on the near-transition regions (rows 2 and 4), we observe an
enhanced migration activity with respect to that found for both
hemispheres (rows 1 and 5). This nicely confirms the locality
assumption, $\Delta y=\Delta\tilde y+\epsilon$.

\paragraph{Scale variations.}

We have checked the scale dependence of the \Sherpa\ \Css\
predictions, which we illustrate for $A_{\rm FB}(p_{T,t\bar t})$ in
the right panel of \fig{fig:pTtt:scales}.
The impact of using the default, $m_{\perp,t}$-like scale choice
($m^2_{\perp,t}=m^2_t+p^2_{T,t}$) versus a fixed $m_t$ scale, and
varying the renormalization and factorization scales simultaneously in
each case by factors of 2, is seen to be of the order of $\pm20$\% at
intermediate $p_{T,t\bar t}$.  This order of magnitude is consistent
with that expected for a leading-order quantity, from variation of the
scales and PDFs in the matrix element and parton showers, and is
generally {\em less}\/ than the variation due to different ways of
treating recoils, or assessing the $p_{T,t\bar t}$ distributions
themselves.

\bibliography{QCDtasy.bib}
\bibliographystyle{utphys}

\end{document}